\documentclass[preprint,prd,nofootinbib,tightenlines,groupedaddress,amsmath,amssymb]{revtex4}

\usepackage{bm}% bold math
\usepackage{color}
\usepackage{graphicx}% Include figure files
\usepackage[hypertex]{hyperref}
\usepackage{multirow}

\begin{document}
\baselineskip=15pt \parskip=3pt

\vspace*{3em}

\title{Lepton-Flavored Scalar Dark Matter with Minimal Flavor Violation}

\author{Chao-Jung Lee and Jusak Tandean}
\affiliation{Department of Physics and Center for Theoretical  Sciences,
National Taiwan University, \\ Taipei 106, Taiwan \\
$\vphantom{\bigg|_{\bigg|}^|}$}

%\date{\today $\vphantom{\bigg|_{\bigg|}^|}$}

\begin{abstract}
We explore scalar dark matter that is part of a lepton flavor triplet satisfying
symmetry requirements under the hypothesis of minimal flavor violation.
Beyond the standard model, the theory contains in addition three right-handed neutrinos
that participate in the seesaw mechanism for light neutrino mass generation.
The dark-matter candidate couples to standard-model particles via Higgs-portal renormalizable
interactions as well as to leptons through dimension-six operators, all of which have minimal
flavor violation built-in.
We consider restrictions on the new scalars from the Higgs boson measurements, observed relic
density, dark-matter direct detection experiments, LEP\,\,II measurements on $e^+e^-$
scattering into a photon plus missing energy, and searches for flavor-violating lepton decays.
The viable parameter space can be tested further with future data.
Also, we investigate the possibility of the new scalars' couplings accounting for the tentative
hint of Higgs flavor-violating decay \,$h\to\mu\tau$\, recently detected in the CMS
experiment.
They are allowed by constraints from other Higgs data to produce a~rate of this decay roughly
compatible with the CMS finding.
\end{abstract}

\maketitle

\section{Introduction\label{intro}}

It is now widely accepted that dark matter (DM) exists in the Universe.
Many observations have led to the inference that DM makes up almost 27\%
of the cosmic energy density budget~\cite{pdg}.
In spite of the evidence, however, the identity of the basic constituents of DM
has continued to be elusive, with the data suggesting that new physics
beyond the standard model (SM) is needed to account for it~\cite{dmreview}.

The necessity for invoking new physics is even more obvious in the treatment of neutrinos.
Since they stay massless in the SM, it cannot explain the numerous measurements
of nonzero neutrino mass and mixing~\cite{pdg}.
Another longstanding and related conundrum is whether neutrinos are Dirac or Majorana
particles.

In the absence of clear empirical guidance about how to address these problems, it is of
interest to entertain various possibilities.
Among the most appealing are models that link the DM and neutrino sectors in such a way
that solves the puzzles in an interconnected or unified manner.
In this paper, we explore a~scenario along a similar line, where DM carries lepton-flavor
quantum numbers and its interactions have some linkage to what makes neutrinos massive.
To make the neutrino connection, we adopt the framework of so-called minimal flavor
violation (MFV).

Motivated by the fact that the SM has been very successful in describing the existing data
on flavor-changing neutral currents and $CP$-violating processes in the quark sector,
the MFV hypothesis postulates that Yukawa couplings are the only sources for the breaking
of flavor and $CP$ symmetries~\cite{mfv1,D'Ambrosio:2002ex,Cirigliano:2005ck}.
Its application to the study of DM carrying quark-flavor quantum numbers was first proposed
in Ref.\,\cite{Batell:2011tc}.
The stability of the quark-flavored DM is due to the presence of an accidental
discrete symmetry which is an element of the combined color and quark-flavor group
under the MFV assumption~\cite{Batell:2013zwa}.

Although the implementation of MFV for quarks is straightforward, there is no unique way to
extend the notion of MFV to the lepton sector, as the SM by itself does not accommodate
lepton-flavor violation.
Since significant flavor mixing among neutrinos has been measured, it is interesting to
formulate MFV for leptons by incorporating ingredients beyond the SM that can account
for this observation~\cite{Cirigliano:2005ck}.
Thus, here we consider the SM slightly expanded with the addition of three
right-handed neutrinos plus a lepton-flavor triplet of scalar fields which has transformation
properties satisfying the MFV principle and contains DM of the popular weakly interacting
massive particle (WIMP) type.
The right-handed neutrinos allow us to activate the usual type-I seesaw mechanism which
results in Majorana neutrinos with small masses~\cite{seesaw1}.
We will not focus on the less interesting possibility of Dirac neutrinos.
Another difference from the quark case is that MFV does not in general lead to longevity for
lepton-flavored DM because of lack of a counterpart of the accidental symmetry which keeps
quark-flavored DM stable~\cite{Batell:2013zwa}.
Therefore, to ensure the stability of our DM candidate we impose a $Z_2$ symmetry under
which the triplet scalars are odd and other particles even.

In the next section, we briefly review the MFV framework in the lepton sector.
In Section$\;$\ref{lfdm}, we describe the Lagrangians with MFV built-in for the scalar triplet.
We assign its quantum numbers in analogy to its quark-flavor counterpart
discussed in the literature~\cite{Batell:2011tc,Lopez-Honorez:2013wla}.
Accordingly, the triplet can interact with SM particles via a Higgs-boson portal at
the renormalizable level and also couple to SM leptons through effective dimension-six
operators.
Section$\;$\ref{numerics} contains our numerical analysis.
We explore constraints on the two types of DM-SM interactions from the Higgs boson data,
observed relic abundance, DM direct detection experiments, LEP\,\,II measurements
of $e^+e^-$ collisions into a photon plus missing energy, and searches for flavor-violating
charged lepton decays.
In addition, we examine whether the new scalars' interactions can explain the recent potential
indication from the CMS experiment of the Higgs flavor-violating decay \,$h\to\mu\tau$\, which
would be an unmistakable signal of physics beyond the SM if confirmed by future measurements.
We make our conclusions in Section$\;$\ref{conclusion}.
Some lengthy formulas are relegated to a~few appendices.

\section{Minimal lepton flavor violation framework\label{mfv}}

In the SM supplemented with three right-handed neutrinos, the renormalizable Lagrangian
for lepton masses can be written as
\begin{eqnarray} \label{Lm}
{\cal L}_{\rm m}^{} \,\,=\,\, -(Y_\nu)_{kl}^{}\,\bar L_{k,L\,}^{}\nu_{l,R\,}^{}\tilde H
\,-\, (Y_e)_{kl}^{}\,\bar L_{k,L\,}^{}E_{l,R\,}^{}H  \,-\,
\mbox{$\frac{1}{2}$}(M_\nu)_{kl}^{}\,\overline{\nu^{\rm c}}_{\!\!\!k,R}^{}\,\nu_{l,R}^{}
\;+\; {\rm H.c.} ~,
\end{eqnarray}
where \,$k,l=1,2,3$\, are summed over, $L_{k,L}$ represents left-handed lepton
doublets, $\nu_{l,R}^{}$ $(E_{l,R})$ denotes right-handed neutrinos (charged leptons),
$Y_{\nu,e}$ are matrices for the Yukawa couplings, $H$~is the Higgs doublet,
\,$\tilde H=i\tau_2^{}H^*$,\, and $M_\nu$~is the Majorana mass matrix for~$\nu_{l,R}^{}$.
The $M_\nu$ part is essential for the type-I seesaw mechanism to generate light
neutrino masses~\cite{seesaw1}.

If neutrinos are Dirac fermions, the $M_\nu$ terms are absent from Eq.\,(\ref{Lm}), and
the MFV hypothesis~\cite{Cirigliano:2005ck} then implies that ${\cal L}_{\rm m}$ has formal
invariance under the~global group
\,${\rm U}(3)_L\times{\rm U}(3)_\nu\times{\rm U}(3)_E =
G_\ell\times{\rm U}(1)_L\times{\rm U}(1)_\nu\times{\rm U}(1)_E$,\,
with \,$G_\ell={\rm SU}(3)_L\times{\rm SU}(3)_\nu\times{\rm SU}(3)_E$\, being the flavor symmetry.
This entails that $L_{k,L}$, $\nu_{k,R}$, and $E_{k,R}$ transform as
fundamental representations of SU$(3)_{L,\nu,E}$, respectively,
\begin{eqnarray}
L_L^{} \,\to\, V_L^{}L_L^{} \,, ~~~~~~~ \nu_R^{} \,\to\, V_\nu^{}\nu_R^{} \,, ~~~~~~~
E_R^{} \,\to\, V_E^{}E_R^{} \,, ~~~~~~~ V_{L,\nu,E}^{}\,\,\in\,\,{\rm SU(3)}_{L,\nu,E}^{} \,,
\end{eqnarray}
whereas the Yukawa couplings transform in the spurion sense according to
\begin{eqnarray}
Y_\nu^{} \,\to\, V_L^{}Y_\nu^{}V^\dagger_\nu \,, ~~~~~~~
Y_e^{} \,\to\, V_L^{}Y_e^{}V^\dagger_E \,.
\end{eqnarray}

Taking advantage of the symmetry under $G_\ell$, we work in the basis where
\begin{eqnarray}
Y_e \,\,=\,\, \frac{\sqrt2}{v}\, {\rm diag}\bigl(m_e^{},m_\mu^{},m_\tau^{}\bigr) \,,
\end{eqnarray}
with \,$v\simeq246$\,GeV\, being the vacuum expectation value of $H$,
and the fields $\nu_{k,L}$, $\nu_{k,R}$, $E_{k,L}$, and $E_{k,R}$ refer to the mass eigenstates.
We can then express $L_{k,L}$ and $Y_\nu$ in terms of the Pontecorvo-Maki-Nakagawa-Sakata
(PMNS~\cite{pmns})
neutrino mixing matrix $U_{\scriptscriptstyle\rm PMNS}$~as
\begin{eqnarray}
L_{k,L}^{} \,= \left( \!\begin{array}{c} (U_{\scriptscriptstyle\rm PMNS})_{kl\,}^{}
\nu_{l,L}^{} \vspace{2pt} \\ E_{k,L}^{} \end{array}\! \right) , ~~~~~~~
Y_\nu \,=\, \frac{\sqrt2}{v}\,U_{\scriptscriptstyle\rm PMNS}^{}\,\hat m_\nu^{} \,, ~~~~
\hat m_\nu^{} \,=\, {\rm diag}\bigl(m_1^{},m_2^{},m_3^{}\bigr) \,, \label{Ynud}
\end{eqnarray}
where $m_{1,2,3}^{}$ are the light neutrino eigenmasses and
in the standard parametrization~\cite{pdg}
\begin{eqnarray} \label{pmns}
U_{\scriptscriptstyle\rm PMNS}^{} \,= \left(\!\begin{array}{ccc}
 c_{12\,}^{}c_{13}^{} & s_{12\,}^{}c_{13}^{} & s_{13}^{}\,e^{-i\delta}
\vspace{1pt} \\
-s_{12\,}^{}c_{23}^{}-c_{12\,}^{}s_{23\,}^{}s_{13}^{}\,e^{i\delta} & ~~
 c_{12\,}^{}c_{23}^{}-s_{12\,}^{}s_{23\,}^{}s_{13}^{}\,e^{i\delta} ~~ & s_{23\,}^{}c_{13}^{}
\vspace{1pt} \\
 s_{12\,}^{}s_{23}^{}-c_{12\,}^{}c_{23\,}^{}s_{13}^{}\,e^{i\delta} &
-c_{12\,}^{}s_{23}^{}-s_{12\,}^{}c_{23\,}^{}s_{13}^{}\,e^{i\delta} & c_{23\,}^{}c_{13}^{}
\end{array}\right) ,
\end{eqnarray}
with $\delta$ being the $CP$ violation phase, \,$c_{kl}^{}=\cos\theta_{kl}^{}$,\, and
\,$s_{kl}^{}=\sin\theta_{kl}^{}$.\,

\pagebreak

If neutrinos are Majorana in nature, $Y_\nu$ must be modified.
The presence of $M_\nu$ in Eq.\,(\ref{Lm}) with nonzero elements much bigger than those of
\,$v Y_\nu/\sqrt2$\, activates the seesaw mechanism~\cite{seesaw1},
leading to the light neutrinos' mass matrix
\begin{eqnarray} \label{mnu}
m_\nu \,\,=\,\, -\frac{v^2}{2}\, Y_\nu^{}M_\nu^{-1}Y_\nu^{\rm T} \,\,=\,\,
U_{\scriptscriptstyle\rm PMNS\,}^{}\hat m_{\nu\,}^{}U_{\scriptscriptstyle\rm PMNS}^{\rm T} \,,
\end{eqnarray}
where now $U_{\scriptscriptstyle\rm PMNS}$ contains the diagonal matrix
\,$P={\rm diag}(e^{i\alpha_1/2},e^{i\alpha_2/2},1)$\, multiplied from the right and
involving the Majorana phases $\alpha_{1,2}^{}$.
This allows one to write~\cite{Casas:2001sr}
\begin{eqnarray} \label{Ynum}
Y_\nu^{} \,\,=\,\,
\frac{i\sqrt2}{v}\,U_{\scriptscriptstyle\rm PMNS\,}^{}\hat m^{1/2}_\nu OM_\nu^{1/2} \,,
\end{eqnarray}
where $O$ is in general a complex matrix satisfying \,$OO^{\rm T}=\openone$,\, the right-hand
side being a 3$\times$3 unit matrix, and \,$M_\nu={\rm diag}(M_1,M_2,M_3)$.\,
From this point on, we assume that neutrinos are Majorana particles and
entertain the possibility that the right-handed neutrinos are degenerate, so that
\,$M_\nu={\cal M}\openone$\, with $\cal M$ being their mass.
In this scenario, the $M_\nu$ part of ${\cal L}_{\rm m}$ breaks SU(3)$_\nu$ into
O(3)$_\nu$, and as a consequence we have \,$G_\ell={\cal G}_\ell\times{\rm O}(3)_\nu$,\,
where \,${\cal G}_\ell={\rm SU}(3)_L\times{\rm SU}(3)_E$\, is the pertinent flavor group
after the heavy right-handed neutrinos are integrated out\,\,\cite{Cirigliano:2005ck}.

To put together Lagrangians beyond the SM with MFV built-in, one inserts $Y_{\nu,e}$,
$Y_{\nu,e}^\dagger$, and their products among SM and new fields to construct
${\cal G}_\ell$-invariant operators that are singlet
under the SM gauge group~\cite{D'Ambrosio:2002ex,Cirigliano:2005ck}.
Of interest here are the matrix products \,${\sf A}=Y_\nu^{}Y_\nu^\dagger$\, and
\,${\sf B}=Y_e^{}Y^\dagger_e$,\, which transform as $(1\oplus8,1)$ under\,\,${\cal G}_\ell$,
as $Y_\nu$ and $Y_e$ transform as $(3,1)$ and $(3,\bar 3)$, respectively.
In a model-independent approach, combinations of $\sf A$ and $\sf B$ are collected into
an object $\Delta$ which formally comprises an infinite number of terms, namely
\,$\Delta=\sum\xi_{jkl\cdots\,}^{}{\sf A}^j{\sf B}^k{\sf A}^l\cdots$\, with coefficients
$\xi_{jkl\cdots}^{}$ expected to be at most of~${\cal O}$(1).
Under the MFV hypothesis, $\xi_{jkl...}^{}$ are real because complex $\xi_{jkl...}^{}$
would introduce new $CP$-violation sources beyond that in the Yukawa couplings.
With the Cayley-Hamilton identity
\,$X^3=X^2\,{\rm Tr}X+\frac{1}{2}X\bigl[{\rm Tr}X^2-({\rm Tr}X)^2\bigr]+\openone{\rm Det}X$\,
for an invertible 3$\times$3 matrix $X$, one can resum the infinite series into a finite
number of terms~\cite{Colangelo:2008qp}:
\begin{eqnarray} \label{Delta0}
\Delta &\,=\,& \xi^{}_1\openone + \xi^{}_{2\,}{\sf A}+\xi^{}_{3\,}{\sf B}
+ \xi^{}_{4\,}{\sf A}^2 +\xi^{}_{5\,}{\sf B}^2 + \xi^{}_{6\,}{\sf AB} + \xi^{}_{7\,}{\sf BA}
+ \xi^{}_{8\,}{\sf ABA} + \xi^{}_{9\,}{\sf BA}^2 + \xi^{}_{10\,}{\sf BAB}
+ \xi^{}_{11\,}{\sf AB}^2
\nonumber \\ && \! +~
\xi^{}_{12\,}{\sf ABA}^2 + \xi^{}_{13\,}{\sf A}^2{\sf B}^2
+ \xi^{}_{14\,}{\sf B}^2{\sf A}^2 + \xi^{}_{15\,}{\sf B}^2{\sf AB}
+ \xi^{}_{16\,}{\sf AB}^2{\sf A}^2+\xi^{}_{17\,}{\sf B}^2{\sf A}^2{\sf B} \;.
\end{eqnarray}
Although $\xi_{ijk\cdots}$ are real, the reduction of the infinite series into the 17 terms
can make the coefficients $\xi_r^{}$ in Eq.\,(\ref{Delta0}) complex due to imaginary parts
among the traces of the matrix products~\,${\sf A}^i{\sf B}^j{\sf A}^k\cdots$.\,
Such imaginary contributions turn out to be small~\cite{Colangelo:2008qp,He:2014fva}, and
so hereafter we ignore~${\rm Im}_{\,}\xi_r$.

In the Dirac neutrino case, $Y_\nu$ in Eq.\,(\ref{Ynud}) leads to
\,${\sf A}=2 U_{\scriptscriptstyle\rm PMNS}^{}\hat m_\nu^2
U_{\scriptscriptstyle\rm PMNS}^\dagger/v^2$,\,
which has tiny elements.
In contrast, if neutrinos are of Majorana nature,
\begin{eqnarray} \label{Am}
{\sf A} \,\,=\,\, \frac{2}{v^2}\, U_{\scriptscriptstyle\rm PMNS\,}^{} \hat m^{1/2}_\nu O
M_\nu^{}O^\dagger\hat m^{1/2}_\nu U_{\scriptscriptstyle\rm PMNS}^\dagger
\end{eqnarray}
from Eq.\,(\ref{Ynum}), and so $\sf A$ can have much greater elements if
the right-handed neutrinos' mass $\cal M$ in $M_\nu$ is sufficiently large.
Since as an infinite series $\Delta$ has to converge, $\cal M$ cannot
be arbitrarily large~\cite{Colangelo:2008qp,He:2014fva}.
Accordingly, we require the largest eigenvalue of $\sf A$ to be unity, which
implies that the elements of \,${\sf B}=Y_e^{}Y^\dagger_e$\, are small compared to those
of $\sf A$ and that, consequently, we can drop most of the terms in Eq.\,(\ref{Delta0})
except the first few.
It follows that in this study
\begin{eqnarray} \label{Delta}
\Delta \,\,=\,\, \xi^{}_1\openone + \xi^{}_{2\,}{\sf A} + \xi^{}_{4\,}{\sf A}^2
\,\,=\,\, \Delta^\dagger \,.
\end{eqnarray}

\section{Lepton-flavored dark matter\label{lfdm}}

The new sector of the theory also includes three complex scalar fields which are singlet
under the SM gauge group and constitute a triplet under
\,${\cal G}_\ell ={\rm SU}(3)_L\times{\rm SU}(3)_E$,\, namely\footnote{Lepton
flavor triplets with DM components have also been considered in the contexts of other
models~\cite{Agrawal:2011ze}.\smallskip}
\begin{eqnarray} \label{tildes}
\tilde s \,\,=\, \left(\begin{array}{c} \tilde s_1^{} \\ \tilde s_2^{} \\ \tilde s_3^{}
\end{array}\right)
\,\sim\,\, (3,1) \,.
\end{eqnarray}
To maintain the longevity of its lowest-mass eigenstate as the DM candidate, we invoke
a $Z_2$ symmetry under which $\tilde s$ is odd and other particles are even.\footnote{Outside
the MFV framework, it is possible to have a DM-stabilizing $Z_2$ symmetry that is a remnant
of a lepton flavor group~\cite{discretedm}.}
This will disallow ${\cal G}_\ell$-invariant interaction terms involving odd numbers of
$\tilde s_k^{(*)}$ that could cause the DM state to decay.

It follows that the renormalizable Lagrangian for
the interactions of the scalar fields with one another and the SM gauge bosons is given by
\begin{eqnarray}
{\cal L} &\,=\,& ({\cal D}^\eta H)^\dagger\,{\cal D}_\eta H
+ \partial^\eta\tilde s^\dagger\,\partial_\eta^{}\tilde s \,-\, {\cal V} \,,
\vphantom{|_{\int_|^|}^{}} \\
\label{V}
{\cal V} &\,=\,& \mu_{H\,}^2H^\dagger H \,+\, \tilde s^\dagger\mu_{s\,}^2\tilde s
\,+\, \lambda_{H\,}^{}(H^\dagger H)^2
\,+\, 2_{\,}H^\dagger H\,\tilde s^\dagger\Delta_{HS\,}\tilde s \,+\,
\bigl(\tilde s^\dagger\Delta_{SS\,}\tilde s\bigr)\raisebox{1pt}{$^2$}
\nonumber \\ &\,\supset\,&
\tilde s^\dagger \bigl(\mu_{s0}^{2~}\openone+\mu_{s1}^{2~}{\sf A}+\mu_{s2}^{2\;\;}{\sf A}^2\bigr)
\tilde s \,+\, 2H^\dagger H\,\tilde s^\dagger \bigl( \lambda_{s0}^{~~}\openone
+ \lambda_{s1}^{~~}{\sf A} + \lambda_{s2}^{~~\,}{\sf A}^2 \bigr) \tilde s
\nonumber \\ && \! +\;
\bigl[\tilde s^\dagger \bigl( \lambda_{s0}^{'~}\openone + \lambda_{s1}^{'~}{\sf A}
+ \lambda_{s2}^{'~\,}{\sf A}^2 \bigr) \tilde s \bigr]\raisebox{1pt}{$^2$} \,,
\end{eqnarray}
where ${\cal D}_\eta$ is the covariant derivative involving the gauge fields,
$\mu_s^2$ and $\Delta_{HS,SS}$ are 3$\times$3
matrices, and the Higgs doublet after electroweak symmetry breaking
\begin{eqnarray}
H \,\,=\, \left(\!\begin{array}{c} 0 \\ \frac{1}{\sqrt2}(h+v) \end{array}\!\right) ,
\end{eqnarray}
with $h$ being the physical Higgs field.
The expression for $\mu_s^2$ $(\Delta_{HS,SS})$ has the form in Eq.\,(\ref{Delta}) up to
an overall factor with mass dimension~2~(0), and hence
the parameters $\mu_{sj}^2$,  $\lambda_{sj}^{}$, and $\lambda_{sj}'$ are real.

With $\sf A$ being Hermitian, we have the relation
\,${\sf A}={\cal U}\,{\rm diag}\bigl(\hat{\mbox{\small$\sf A$}}_1,
\hat{\mbox{\small$\sf A$}}_2,\hat{\mbox{\small$\sf A$}}_3\bigr)\,{\cal U}^\dagger$\,
where $\cal U$ is a unitary matrix and
$\hat{\mbox{\small$\sf A$}}_k$ denotes the eigenvalues of $\sf A$.
Accordingly, the matrices sandwiched between $\tilde s^\dagger$ and
$\tilde s$  in Eq.\,(\ref{V}) can be simultaneously diagonalized.
It follows that $\tilde s_k^{}$ are related to the mass eigenstates $S_k$ by
\begin{eqnarray}
S \,\,=\, \left(\begin{array}{c} S_1^{} \\ S_2^{} \\ S_3^{} \end{array}\right) \,=\,\,
{\cal U}^\dagger\tilde s \,,
\end{eqnarray}
in terms of which
\begin{eqnarray} \label{hss}
{\cal L} \,\,\supset\,\, -m_{S_k}^2S_k^*S_k^{}\,-\,\lambda_{k\,}^{}\bigl(h^2+2hv\bigr)S_k^*S_k^{}
\,-\, \bigl(\lambda_{k\,}'S_k^*S_k^{}\bigr)\raisebox{1pt}{$^2$} \,,
\end{eqnarray}
where summation over $k$ is implicit,
\begin{eqnarray} \label{smasses}
m_{S_k}^2 \,=\, \mu_k^2 \,+\, \lambda_k^{}v^2 \,, ~~~~~~~
\mu_k^2 \,=\, \mu_{s0}^2 + \mu_{s1\,}^2\hat{\mbox{\small$\sf A$}}_k^{}
+ \mu_{s2\,}^2\hat{\mbox{\small$\sf A$}}_k^2 \,, ~~~~~~~
\lambda_k^{(\prime)} \,=\, \lambda_{s0}^{(\prime)}
+ \lambda_{s1\,}^{(\prime)}\hat{\mbox{\small$\sf A$}}_k^{}
+ \lambda_{s2\,}^{(\prime)}\hat{\mbox{\small$\sf A$}}_k^2 \,. ~~~~
\end{eqnarray}

\pagebreak

Since $\mu_{si}^2$ and $\lambda_{si}^{(\prime)}$ are free parameters, so are \,$m_{S_k}>0$\,
and $\lambda_k^{(\prime)}$.
There are, however, theoretical restrictions on $\lambda_k^{(\prime)}$ as well as
$\lambda_H$.
The stability of the vacuum requires $\cal V$ to be bounded from below, which entails
\,$\lambda_H^{}>0$,\, $\bigl(\lambda_k'\bigr)\raisebox{1pt}{$^2$}>0$,\, and
\,$\lambda_k^{}>-\sqrt{\lambda_H}\;\bigl|\lambda_k'\bigr|$,\, the second inequality being
automatically satisfied by the reality of $\lambda_k'$.
The condition of perturbativity~\cite{perturbativity} translates into
\,$|\lambda_{H,k}|<4\pi$\, and \,$(\lambda_k')^2<4\pi$.\,

The $\lambda_k$ part in Eq.\,(\ref{hss}) is responsible for the Higgs-portal interactions
of the new scalars with SM particles.
As we detail later, in this paper we select $S_3$ to be less massive than $S_{1,2}$ and serve
as the DM candidate.
In addition, we pick the $S_{1,2}$ masses to be sufficiently bigger than $m_{S_3}$ in order
that their impact on the relic density can be ignored.
In that case, $\lambda_3$ controls the Higgs-mediated annihilations of the DM into SM particles,
its scattering off a nucleon via Higgs exchange, and also the Higgs nonstandard invisible decay
if the $S_3$ mass is low enough.
All of these processes are subject to constraints  from various recent data.

Because of their flavor quantum numbers in Eq.\,(\ref{tildes}), the new particles cannot
have renormalizable contact interactions with SM fermions.
Rather, under the MFV framework supplemented with the DM stabilizing $Z_2$ symmetry,
$S_k$ can couple with SM leptons due to effective operators of dimension six given
by\footnote{Without the $Z_2$ symmetry, the DM candidate could undergo rapid decay triggered
by effective operators involving odd numbers of $\tilde s$, such as
\,$\epsilon_{bdk\,}^{}\overline{(\Delta_1L_L)\mbox{$_b^{\rm c}$}}\,\tilde H^*\tilde H^\dagger
(\Delta_2L_L)_d^{}\,(\Delta_3 \tilde s)_k^{}$,\,
where $\Delta_{1,2,3}$ are of the form in Eq.\,(\ref{Delta}) with their respective
coefficients $\xi$'s.\smallskip}
\begin{eqnarray} \label{L'}
{\cal L}' \,\,=\,\, \frac{C_{bdkl}^{\scriptscriptstyle L}}{\Lambda^2}\,
O_{bdkl}^{\scriptscriptstyle L}
+ \frac{C_{bdkl}^{\scriptscriptstyle R}}{\Lambda^2}\,O_{bdkl}^{\scriptscriptstyle R}
+ \biggl(\frac{C_{bdkl}^{\scriptscriptstyle LR}}{\Lambda^2}\,
O_{bdkl}^{\scriptscriptstyle LR}\;+\;{\rm H.c.}\biggr) \,,
\end{eqnarray}
where summation over \,$b,d,k,l=1,2,3$\, is implicit,
\begin{eqnarray} \label{CO}
C_{bdkl}^{\scriptscriptstyle L} &=&
(\Delta_{\scriptscriptstyle LL})_{bd}^{}(\Delta_{\scriptscriptstyle SS})_{kl}^{}
+ (\Delta_{\scriptscriptstyle LS})_{bl}^{}(\Delta_{\scriptscriptstyle SL})_{kd}^{}
+ (\Delta_{\scriptscriptstyle LS})_{kd}^{}(\Delta_{\scriptscriptstyle SL})_{bl}^{} \,, ~~~~
O_{bdkl}^{\scriptscriptstyle L} \,=\, i\bar L_{b,L}^{}\gamma^\rho L_{d,L\,}^{}\tilde s_k^*
\raisebox{1pt}{\small$\stackrel{\scriptscriptstyle\leftrightarrow}{\partial}$}_{\!\rho}^{}
\tilde s_l^{} \,,
\nonumber \\
C_{bdkl}^{\scriptscriptstyle R} &=&
\delta_{bd\,}^{}\bigl(\Delta_{\scriptscriptstyle SS}'\bigr)_{kl} \,, \hspace{216pt}
O_{bdkl}^{\scriptscriptstyle R} \,=\, i\bar E_{b,R}^{}\gamma^\rho E_{d,R\,}^{}\tilde s_k^*
\raisebox{1pt}{\small$\stackrel{\scriptscriptstyle\leftrightarrow}{\partial}$}_{\!\rho}^{}
\tilde s_l^{} \,, \vphantom{|_{\int}^{}}
\nonumber \\
C_{bdkl}^{\scriptscriptstyle LR} &=& (\Delta_{\scriptscriptstyle LY}Y_e)_{bd}^{}
\bigl(\Delta_{\scriptscriptstyle SS}''\bigr)_{kl}
+ \bigl(\Delta_{\scriptscriptstyle LS}'\bigr)_{bl} (\Delta_{\scriptscriptstyle SY}Y_e)_{kd} \,,
\hspace{80pt} O_{bdkl}^{\scriptscriptstyle LR} =\,
\bar L_{b,L}^{}E_{d,R\,}^{}\tilde s_k^*\tilde s_{l\,}^{}H \,, ~~~~~~~~~
\end{eqnarray}
with\footnote{The counterparts of $O^{{\scriptscriptstyle L},\scriptscriptstyle E}$ with
$\tilde s_k^*\raisebox{1pt}{\scriptsize$
\stackrel{\scriptscriptstyle\leftrightarrow}{\partial}$}_{\!\rho}^{}\tilde s_l^{}$
replaced by \,$\tilde s_k^*\,\partial_\rho\tilde s_l^{}+\partial_\rho\tilde s_k^*\,\tilde s_l^{}$\,
are not independent and can be expressed in terms of $O^{{\scriptscriptstyle LR}(\dagger)}$
after partial integration and use of the lepton equations of motion~\cite{Kamenik:2011vy}.}
\,$X\raisebox{1pt}{\footnotesize$\stackrel{\scriptscriptstyle\leftrightarrow}{\partial}$}_{\!\rho}Y
=X_{\,}\partial_\rho Y-\partial_\rho X_{\,} Y$\,
and \,$\tilde s_k^{}={\cal U}_{kl\,}^{}S_l^{}$.\,
We have dropped terms in $C^{\scriptscriptstyle R}$ that are suppressed by two powers of $Y_e$.
Since the right-handed neutrinos have masses far exceeding the TeV level, we do not include
operators involving them in ${\cal L}'$.
The mass scale $\Lambda$ characterizes the heavy new physics underlying these interactions
and also responsible for the Lorentz and flavor structure of the operators.
Specifically, $O^{\scriptscriptstyle L,R}$ ($O^{\scriptscriptstyle LR}$) could arise from
the exchange of a spin-one boson (scalar or fermion), and so
$\Lambda$ would depend on its couplings and mass.

The $\Delta$'s in $C^{\scriptscriptstyle L,R,LR}$ above are of the same form as
in~Eq.\,(\ref{Delta}), but have generally different coefficients $\xi$'s.
These $\xi$'s are expected to be at most of ${\cal O}(1)$, and some of them may be suppressed
or vanish, depending on the underlying theory.
In our model-independent approach with MFV, we single out a~few of them in order to illustrate
some of the phenomenological implications.

\section{Numerical analysis\label{numerics}}

With $S_3^{}$ being the DM, the cross section $\sigma_{\rm ann}^{}$ of $S_3^{}S_3^*$
annihilation into SM particles needs to yield the present-day DM density $\Omega$.
The two quantities are approximately related by~\cite{Kolb:1990vq}
\begin{eqnarray}
\Omega\hat h^2 \,\,=\,\,
\frac{2.14\times10^9\,x_f^{}\,\rm\,GeV^{-1}}{\sqrt{g_*^{}}\,m_{\rm Pl}^{}\,
\bigl(\hat a+3\hat b/x_f^{}\bigr)} ~, ~~~~~~~
x_f^{} \,\,=\,\, \ln\frac{0.038\,m_{S_3}^{}\,m_{\rm Pl}^{}\,
\bigl(\hat a+6\hat b/x_f^{}\bigr)}{\sqrt{g_*^{}\,x_f^{}}} ~,
\end{eqnarray}
where $\hat h$ stands for the Hubble parameter, \,$m_{\rm Pl}^{}=1.22\times10^{19}$\,GeV\, is
the Planck mass, $g_*^{}$ is the number of relativistic degrees of freedom below the freeze-out
temperature~\,$T_f^{}=m_{S_3}/x_f^{}$, and $\hat a$ and $\hat b$ are defined by the expansion
of the annihilation rate \,$\sigma_{\rm ann}^{}v_{\rm rel}^{}=\hat a+\hat bv_{\rm rel}^2$\, in
terms of the relative speed $v_{\rm rel}^{}$ of the nonrelativistic $S_3^{}S_3^*$ pair in
their center-of-mass (c.m.) frame.
The $\Omega$ expression takes into account the fact that the DM is a complex
scalar particle.

\subsection{Higgs-portal interactions\label{higgsportal}}

The $S_3^{}$ contributions to $\sigma_{\rm ann}^{}$ originate mainly from the $\lambda_3$ term
in Eq.\,(\ref{hss}) as well as from the dimension-6 operators in Eq.\,(\ref{L'}).
We consider first the possibility that the latter are absent.
The $\lambda_3$ coupling gives rise to Higgs-mediated $S_3^{}S_3^*$ collisions into
SM particles, just as in the case of the SM-singlet scalar~DM~\cite{Silveira:1985rk,He:2010nt}.
The resulting annihilation rate in the nonrelativistic limit is dominated by its $\hat a$ part,
\begin{eqnarray} \label{sigmavrel}
\sigma_{\rm ann}^{} v_{\rm rel}^{} \,\,\simeq\,\, \hat a \,\,=\,\,
\frac{4\lambda_{3\,}^2 v^2\,m_{S_3}^{-1}\,\mbox{$\sum_i$}\Gamma\bigl(\tilde h\to X_i\bigr)}
{\bigl(4m_{S_3}^2-m_h^2\bigr){}^{^{\scriptstyle2}}+\Gamma^2_h\,m^2_h} ~,
\end{eqnarray}
where $m_h^{}$ is the mass of the Higgs boson, $\Gamma_h^{}$ is its total width
$\Gamma_h^{\scriptscriptstyle\rm SM}$ in the SM plus the rates of the decays
\,$h\to S_k^{}S_k^*$\, to be discussed below, $\tilde h$~is a virtual Higgs boson having
the same couplings as the physical $h$, but with the invariant mass \,$\sqrt s=2m_{S_3}^{}$,
and  \,$\tilde h\to X_i$\, is any kinematically allowed decay mode of~$\tilde h$.
For~\,$m_{S_3}>m_h^{}$,\, the \,$S_3^{}S_3^*\to hh$\, reaction can happen, due to
$s$-, $t$-, and $u$-channel as well as contact diagrams, and hence needs to be
included in $\hat a$.
Numerically, we employ \,$m_h^{}=125.1$\,GeV,\, which reflects the average of the most recent
measurements~\cite{mhx,CMS:2014ega},
and~\,$\Gamma_h^{\scriptscriptstyle\rm SM}=4.08\;$MeV\,~\cite{lhctwiki}.
Once the $\lambda_3$ values which reproduce the observed relic abundance are extracted,
they need to fulfill important restrictions which we now address.

A number of underground experiments have been performed to detect WIMP DM directly by looking
for the recoil energy of nuclei caused by the elastic scattering of a WIMP off a~nucleon,~$N$.
Our process of interest is \,$S_3^{(*)}N\to S_3^{(*)}N$\, which proceeds mainly via Higgs
exchange in the $t$ channel and hence depends on $\lambda_3$ as well.
Its cross section is
\begin{eqnarray} \label{sel}
\sigma_{\rm el}^{} \,\,=\,\, \frac{\lambda_3^2\,g_{NNh\,}^2 m_{N\,}^2 v^2}
{\pi_{\,}\bigl(m_{S_3}^{}+m_N^{}\bigr)\raisebox{1pt}{$^2$}m_h^4}
\end{eqnarray}
in the nonrelativistic limit, where $m_N^{}$ is the nucleon mass and $g_{NNh}^{}$
denotes the Higgs-nucleon effective coupling whose value is within
the range \,$0.0011\le g_{NNh}^{}\le0.0032$\,~\cite{He:2010nt}.
The null result of searches by the LUX experiment~\cite{Akerib:2013tjd}
translates into the strictest limit to date on~$\sigma_{\rm el}^{}$.

If $m_{S_k}$ is less than half of the Higgs mass, the nonstandard decay channel
\,$h\to S_k^{}S_k^*$\, is open.
This leads to the branching ratio
\begin{eqnarray}
{\cal B}\bigl(h\to S^*S\bigr) \,\,=\,\, \frac{\raisebox{3pt}{\footnotesize$\displaystyle\sum_k$}\,
\Gamma_{h\to S_k^*S_k^{}}^{}}{\Gamma_h^{\scriptscriptstyle\rm SM} +
\raisebox{3pt}{\footnotesize$\displaystyle\sum_k$}\,\Gamma_{h\to S_k^*S_k^{}}^{}} \;,
\end{eqnarray}
where the summation is over final states satisfying \,$2m_{S_k}<m_h^{}$\, and from Eq.\,(\ref{hss})
\begin{eqnarray}
\Gamma_{h\to S_k^*S_k^{}}^{} \,\,=\,\, \frac{\lambda_k^{2\,}v^2}{4\pi_{\,}m_h^{}}
\sqrt{1-\frac{4m_{S_k}^2}{m_h^2}} ~.
\end{eqnarray}
The couplings $\lambda_k$ are thus subject to restrictions on the Higgs
invisible or non-SM decay modes from collider data.

\begin{figure}[b]
\includegraphics[width=93mm]{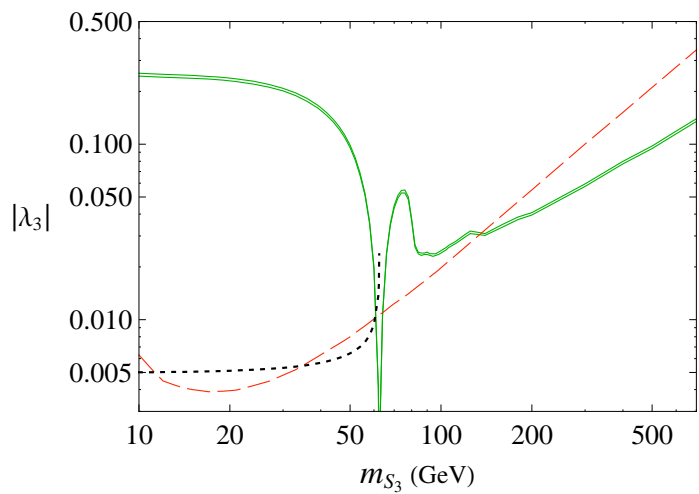}\vspace{-7pt}
\caption{Values of $|\lambda_3|$ consistent with the relic density data (green solid curve),
compared to upper limits on $|\lambda_3|$ from Higgs measurements (black dotted curve) and
from null results of DM direct searches (red dashed curve), as discussed in
the text.\label{lambda3}}
\end{figure}

To determine the $\lambda_3$ values that are consistent with the observed relic density, we
apply the relevant formulas described above and impose \,$0.1155\le\Omega\hat h^2\le0.1241$\,
which is the 90\%\,
confidence level (CL) range of the data \,$\Omega\hat h^2=0.1198\pm0.0026$\,~\cite{pdg}.
In Figure$\;$\ref{lambda3} we display the result (green solid curve) for \,$m_{S_3}\ge10$\,GeV.\,
It needs to be compared with the red dashed curve, which represents the upper limit on
$|\lambda_3|$ inferred from the null result reported by the LUX
Collaboration~\cite{Akerib:2013tjd}.
To arrive at this curve, we used Eq.\,(\ref{sel}) with \,$g_{NNh}^{}=0.0011$,\, which is
the lower end of its range and thus leads to the loosest limit on $|\lambda_3|$ from
the most stringent of DM direct searches to date.
For~\,$2m_{S_3}<m_h^{}$,\, the experimental information on the Higgs nonstandard invisible
decay implies further restraints.
Assuming that the channels \,$h\to S_1^{}S_1^*,S_2^{}S_2^*$\, are absent, we have plotted
the black dotted curve upon demanding \,${\cal B}\bigl(h\to S^*S\bigr)<0.19$\,
based on the bounds from the latest analyses of collider data~\cite{higgsdata}.
The opening of the $S_{1,2}^{}S_{1,2}^*$ channels would cause the dotted curve to shift down.

From the figure, one can infer that the $\lambda_3$ contribution to the annihilation rate
is much less than half of the required amount if \,$m_{S_3}<90$\,GeV,\,
except the neighborhood of \,$m_{S_3}=m_h^{}/2$.\,
In other words, over most of this mass region the $\lambda_3$ term in Eq.\,(\ref{hss})
cannot play the leading role responsible for the observed relic abundance.
Therefore, the dominant contribution must come from the effective interactions
in~Eq.\,(\ref{L'}), absent other DM candidates.
For larger $m_{S_3}$, on the other hand, each of the two sources can generate a nonnegligible
effect on the relic density.

\subsection{Effective DM-lepton interactions\label{SSll}}

The effective operators in Eq.\,(\ref{L'}) induce DM annihilations into SM leptons and are
subject to constraints which may not apply to the Higgs-$S_k$ renormalizable couplings.
From Eq.\,(\ref{CO}), we derive the amplitudes for
the DM annihilation \,$S_3^{}(p)\,S_3^*(\bar p)\to\ell_b^-\ell_d^+,\nu_b^{}\nu_d^{}$\, to be
\begin{eqnarray}
{\cal M}_{S_3^{}\bar S_3^{}\to\ell_b^{}\bar\ell_d^{}}^{} &\,=\,& \frac{1}{\Lambda^2}\,\bar u_b^{}
\biggl[ \frac{-v}{\sqrt2} \bigl( {\cal C}_{bd}^{\scriptscriptstyle LR}P_R^{} +
{\cal C}_{db}^{{\scriptscriptstyle LR}*}P_L^{}\bigr) \,+\,
\gamma^\rho(\bar p-p)_{\rho\,}^{}\bigl({\cal C}_{bd}^{\scriptscriptstyle L}P_L^{}
+ {\cal C}_{bd}^{\scriptscriptstyle R}P_R^{}\bigr) \biggr] v_d^{} \;, ~~~~
\nonumber \\
{\cal M}_{S_3^{}\bar S_3^{}\to\nu_b^{}\nu_d^{}}^{} &\,=\,& \frac{1}{\Lambda^2}(\bar p-p)_\rho^{}\,
\bar u_{b\,}^{}\gamma^\rho \bigl({\cal C}_{bd}^{\scriptscriptstyle L}P_L^{}
- {\cal C}_{db}^{\scriptscriptstyle L}P_R^{}\bigr) v_d^{} \;,
\end{eqnarray}
where $u_b^{}$ and $v_d^{}$ are the leptons' spinors,
\,$P_{L,R}=\frac{1}{2}\bigl(1\mp\gamma_5^{}\bigr)$,\,
\begin{eqnarray} \label{Cbd}
{\cal C}_{bd}^\epsilon \,\,=\,\, \raisebox{3pt}{\footnotesize$\displaystyle\sum_{k,l}$}\;
{\cal U}_{k3}^*\,{\cal U}_{l3}^{}\,C_{bdkl}^\epsilon \;, ~~~~~~~
\epsilon \,\,=\,\, LR,L,R \;,
\end{eqnarray}
and for the $\nu_b^{}\nu_d^{}$ channel we have taken into account the neutrinos' Majorana nature.
The contributions of these reactions to the annihilation rate
\,$\sigma_{\rm ann}^{}v_{\rm rel}^{}=\hat a+\hat b v_{\rm rel}^2$\, are
\begin{eqnarray}
\hat a &\,=\,& \frac{{\cal K}^{\frac{1}{2}}\bigl(4m_{S_3}^2,m_{\ell_o}^2,m_{\ell_r}^2\bigr)_{\,}v^2}
{256\Lambda_{\;\;}^4\pi_{\,}m_{S_3}^4} \Bigl[ \bigl(|{\cal C}_{or}^{\scriptscriptstyle LR}|^2
+ |{\cal C}_{ro}^{\scriptscriptstyle LR}|^2\bigr)\bigl(4m_{S_3}^2-m_{\ell_o}^2-m_{\ell_r}^2\bigr)
\,-\, 4\,{\rm Re}\bigl({\cal C}_{or}^{\scriptscriptstyle LR\,}
{\cal C}_{ro}^{\scriptscriptstyle LR}\bigr)\, m_{\ell_o}^{}m_{\ell_r}^{} \Bigr] \,,
\nonumber \\
\hat b &\,=\,& \frac{{\cal K}^{\frac{1}{2}}\bigl(4m_{S_3}^2,m_{\ell_o}^2,m_{\ell_r}^2\bigr)}
{1536\Lambda_{\;\;}^4\pi_{\,}m_{S_3}^4} \Bigl\{ \bigl(|{\cal C}_{or}^{\scriptscriptstyle L}|^2
+ |{\cal C}_{or}^{\scriptscriptstyle R}|^2\bigr) \Bigl[ 32 m_{S_3}^4
- 4\bigl(m_{\ell_o}^2+m_{\ell_r}^2\bigr)m_{S_3}^2
- \bigl(m_{\ell_o}^2-m_{\ell_r}^2\bigr)\raisebox{1pt}{$^2$} \Bigr]
\nonumber \\ && \hspace{21ex} +\;
48\,{\rm Re}\bigl({\cal C}_{or}^{{\scriptscriptstyle L}*\,}
{\cal C}_{or}^{\scriptscriptstyle R}\bigr)\,m_{S_3}^2 m_{\ell_o}^{}m_{\ell_r}^{} \Bigr\}
\nonumber \\ && \! +\;
\frac{|{\cal C}_{or}^{\scriptscriptstyle L}|^2 m_{S_3}^2}{12\Lambda_{\;\;}^4\pi}
\;+\; ({\cal C}_{or,ro}^{\scriptscriptstyle LR}\rm~terms) \;, \label{hb}
\end{eqnarray}
where \,${\cal K}(x,y,z)=x^2+y^2+z^2-2(x y+y z+x z)$\, and summation over \,$o,r=1,2,3$\, is
implicit, to include all the final lepton states.
For \,$2m_{S_3}>m_{\ell_b}+m_{\ell_d}+m_h^{}$, the \,$\epsilon=LR$\, operator also yields
\,$S_3^{}S_3^*\to\ell_b^-\ell_d^+h$,\, but its impact can be neglected in our $m_{S_3}$ range
of interest.

Since the $\Delta$'s in Eq.\,(\ref{CO}) contain many free parameters, to proceed we need to
make more specific choices regarding $C_{bdkl}^{\scriptscriptstyle LR,L,R}$.
For simplicity, we adopt
\begin{eqnarray} \label{ccc}
C_{bdkl}^{\scriptscriptstyle LR} \,\,=\,\,
\frac{\sqrt2\,\kappa_{LR\,}^{}m_{\ell_d}^{}}{v}\,\delta_{bl}^{}\delta_{dk}^{} \;, ~~~~~~~
C_{bdkl}^{\scriptscriptstyle L} \,\,=\,\, 2\kappa_L^{}\delta_{bl\,}^{}\delta_{dk}^{} \;,
~~~~~~~ 
C_{bdkl}^{\scriptscriptstyle R} \,\,=\,\, \kappa_{R\,}^{}\delta_{bd\,}^{}\delta_{kl}^{} \;,
\end{eqnarray}
with $\kappa_{LR,L,R}^{}$ being real constants.
From Eq.\,(\ref{Cbd}) and the unitarity of $\,\cal U$,  we then have
\begin{eqnarray} \label{CCC}
{\cal C}_{bd}^{\scriptscriptstyle LR} \,\,=\,\, \frac{\sqrt2\,\kappa_{LR\,}^{}m_{\ell_d}^{}}{v}\;
{\cal U}_{b3\,}^{}{\cal U}_{d3}^* \;, ~~~~~~~
{\cal C}_{bd}^{\scriptscriptstyle L} \,\,=\,\, 2 \kappa_L^{}\,
{\cal U}_{b3\,}^{}{\cal U}_{d3}^* \;, ~~~~~~~
{\cal C}_{bd}^{\scriptscriptstyle R} \,\,=\,\, \kappa_{R\,}^{}\delta_{bd}^{} \;.
\end{eqnarray}

We also need to specify the $S_k$ masses.
Among the different ways to realize $\sf A$ in Eq.\,(\ref{Am}), we concentrate on
the least complicated possibility that $O$ is a~real orthogonal matrix, in addition to
the right-handed neutrinos being degenerate with \,$M_\nu={\cal M}\openone$,\, in which case
\begin{eqnarray} \label{Achoice}
{\sf A} \,\,=\,\, \frac{2_{\,}\cal M}{v^2}\,U_{\scriptscriptstyle\rm PMNS\,}^{}
\hat m_{\nu\,}^{}U_{\scriptscriptstyle\rm PMNS}^\dagger \;.
\end{eqnarray}
With \,${\sf A}={\cal U}\,{\rm diag}\bigl(\hat{\mbox{\small$\sf A$}}_1,
\hat{\mbox{\small$\sf A$}}_2,\hat{\mbox{\small$\sf A$}}_3\bigr)\,{\cal U}^\dagger$,\,
this implies that
\begin{eqnarray}
{\cal U} \,\,=\,\, U_{\scriptscriptstyle\rm PMNS}^{} \,, ~~~~~~~
\hat{\mbox{\small$\sf A$}}_k^{} \,\,=\,\, \frac{2{\cal M}_{\,}m_k^{}}{v^2} \,.
\end{eqnarray}
The $S_k$ mass formula in Eq.\,(\ref{smasses}) then becomes
\begin{eqnarray} \label{mSk2}
m_{S_k}^2 \,=\, \mu_{s0}^2+\lambda_{s0}^{}v^2 \,+\,
\frac{2\bigl(\mu_{s1}^2+\lambda_{s1}^{}v^2\bigr){\cal M}_{\,} m_k^{}}{v^2}
\,+\, \frac{4\bigl(\mu_{s2}^2+\lambda_{s2}^{}v^2\bigr){\cal M}_{\,}^2 m_k^2}{v^4} \,,
\end{eqnarray}
indicating that the pattern of $S_k$ masses is connected to the mass hierarchy of the light
neutrinos.
For definiteness, we pick
\begin{eqnarray} \label{mulambda}
\mu_{s0}^2+\lambda_{s0}^{}v^2 \,=\, \mu_{s1}^2+\lambda_{s1}^{}v^2 \,=\,
\mu_{s2}^2+\lambda_{s2}^{}v^2 \,.
\end{eqnarray}
Thus a normal hierarchy of neutrino masses, \,$m_1^{}<m_2^{}\ll m_3^{}$,\, would cause
$S_{1,2}$ to be close in mass and lighter than $S_3$, implying that at least both $S_{1,2}$
determine the DM density.
As stated earlier, here we examine the simpler scenario with the inverted hierarchy of
neutrino masses, \,$m_3^{}\ll m_1^{}<m_2^{}$,\, so that only $S_3$ is the DM and the heavier
$S_{1,2}$ have negligible effects on the relic abundance.

For numerical computations below, we need to know the elements of $\cal U$ as well as
the light neutrino eigenmasses.
We employ the central values of the parameter ranges
\begin{eqnarray}
\sin^2\theta_{12}^{} &\,=\,& 0.308\pm0.017 ~, ~~~~~~~
\sin^2\theta_{23}^{} ~=~ 0.455_{-0.031}^{+0.139} ~,
\nonumber \\
\sin^2\theta_{13}^{} &\,=\,& 0.0240_{-0.0022}^{+0.0019} ~, \hspace{10ex}
\delta/\pi ~=~ 1.31_{-0.33}^{+0.29} ~,
\nonumber \\
\delta m^2 &\,=\,& m_2^2-m_1^2 \,=\,
\left(7.54_{-0.22}^{+0.26}\right)\times10^{-5}\;{\rm eV}^2 ~,
\nonumber \\
\Delta m^2 &\,=\,& \mbox{$\frac{1}{2}$}\bigl(m_1^2+m_2^2\bigr)-m_3^2 \,=\,
\bigl(2.38_{-0.06}^{+0.06}\bigr)\times10^{-3}\;{\rm eV}^2 \label{nudata}
\end{eqnarray}
from a recent fit to the global data on neutrino oscillation~\cite{Capozzi:2013csa}
in the case of inverted hierarchy of neutrino masses.
Since empirical information on the absolute scale of $m_{1,2,3}^{}$ is still far from
precise~\cite{pdg}, we set \,$m_3^{}=0$.\,
Requiring the largest eigenvalue of $\sf A$ in Eq.\,(\ref{Achoice}) to be unity,
we then get \,${\cal M}=6.15\times10^{14}$\,GeV.\,
Applying these mass numbers and Eq.\,(\ref{mulambda}) in Eq.\,(\ref{mSk2}) results in
\,$m_{S_1}\simeq1.7_{\,}m_{S_3\,}$ and $m_{S_1,S_2}$ differing by {\small$\sim$\,}0.8\%.

\begin{figure}[t]
\includegraphics[width=93mm]{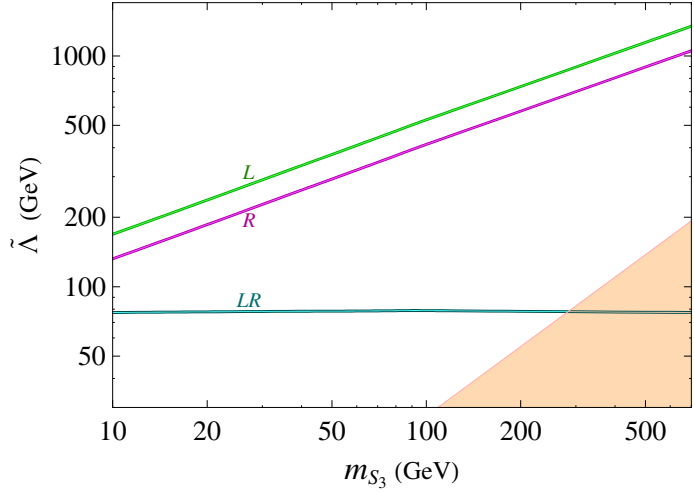}\vspace{-7pt}
\caption{Values of \,$\tilde\Lambda=\Lambda/|\kappa_\epsilon|^{1/2}$\, for \,$\epsilon=L,R,LR$\,
which fulfill the relic density constraint, as discussed in the text.
In this and the following figures, the (orange) shaded region depicts the parameter space
where the effective field theory approach breaks down.\label{tLambda}}
\end{figure}

We can now extract the values of \,$\tilde\Lambda\equiv\Lambda/|\kappa_\epsilon^{}|^{1/2}$\,
that fulfill the relic density requirement using Eq.\,(\ref{hb}) with the couplings
given in Eq.\,(\ref{CCC}).
Assuming that only one of $\kappa_{LR,L,R}$ is nonzero at a~time and that the $\lambda_3$
contributions evaluated earlier are absent, we present the results in Figure$\;$\ref{tLambda}.
The curve for \,$\epsilon=LR$\, arises from $\hat a$ in Eq.\,(\ref{hb}),
with the contribution from $\hat b$ having been neglected, whereas the \,$\epsilon=L$ or $R$\,
curve comes from $\hat b$ alone.\footnote{The roughly flat behavior of the $LR$ (blue) curve
reflects the $m_{S_3}$ independence of $\hat a$ in Eq.\,(\ref{hb}) for negligible lepton masses
and is similar to its counterpart in the quark-flavored-DM scenario~\cite{Batell:2011tc}.}
If the $\lambda_3$ contributions are also present and nonnegligible, and if
they do not cancel the effective-coupling contributions in the \,$\ell_k^-\ell_k^+$\,
channels, there will be less room for each of the two sources,
which will push the $\tilde\Lambda$ curves upward.
On the lower right portion of the plot, we have also drawn an orange area, which satisfies
\,$2\pi\tilde\Lambda<m_{S_2}$\, for the parameter choices in Eq.\,(\ref{mulambda}) and
the preceding paragraph.
This region corresponds to the parameter space where the effective field
theory description is no longer valid~\cite{Goodman:2010ku}.

There is another restraint from DM data that should be mentioned.
For \,$m_{S_3}<20$\,GeV,\, the predicted annihilation rate for the $\tau^+\tau^-$ final-state
is in some tension with upper limits inferred from searches for DM signals in
diffuse gamma-ray data from the Fermi Large Area Telescope observations of dwarf
spheroidal satellite galaxies of the Milky Way~\cite{Ackermann:2011wa}.

More significantly, complementary constraints on the effective lepton-$S$ couplings are
available from experimental studies at LEP\,\,II on the monophoton production process
\,$e^+e^-\to\gamma\,\slash\!\!\!\!E$\, with missing energy \,$\slash\!\!\!\!E$\, in the final state.
These measurements were carried out to examine the neutrino counting reaction
\,$e^+e^-\to\gamma\nu\bar\nu$\, in the SM and also to look for new particles that are long-lived or
stable~\cite{lep:ee2gnn}.
Thus the acquired data may be useful for restricting the process
\,$e^+e^-\to\gamma S_k^{}S_l^*$\, if $S_{k,l}$ are long-lived or, for \,$k,l\neq3$,\, if they decay
(sequentially) into $S_3$ plus light neutrinos.
This transition arises from two diagrams each containing an \,$e^+e^-\to S_k^{}S_l^*$\, vertex with
the photon being radiated off the $e^-$ or $e^+$ line.
We have written down its amplitude and sketched the calculation of the cross section,
$\sigma_{e\bar e\to \gamma S_k\bar S_l}$, in Appendix\,\,\ref{csr}.
Summing it over the final flavors then yields
$\sigma_{e\bar e\to\gamma S\bar S'\to\gamma\,\slash\!\!\!\!E}$ if $S_{k,l}$ are stable or
long-lived.
If they decay, we can express instead
\begin{eqnarray} \displaystyle \label{cs_ee2gSS}
\sigma_{e\bar e\to\gamma S\bar S'\to\gamma\,\slash\!\!\!\!E}^{} \,\,=\,\,
\raisebox{2pt}{\footnotesize$\displaystyle\sum_{k,l\,=1}^3$}\,
\sigma_{e\bar e\to\gamma S_k\bar S_l}^{}\,{\cal B}_{k3}^{}\,{\cal B}_{l3}^{}
\end{eqnarray}
with the branching ratios
\begin{eqnarray}
{\cal B}_{13} \,=\, {\cal B}(S_1\to\nu\nu'S_3) \;, \hspace{4ex}
{\cal B}_{23} \,=\, {\cal B}(S_2\to\nu\nu'S_3)
+ {\cal B}(S_2\to\nu\nu'S_1)\,{\cal B}_{13} \;, \hspace{4ex}
{\cal B}_{33} \,=\, 1 \;, ~~~
\end{eqnarray}
where the sum includes only kinematically allowed channels and
\,${\cal B}(S_l\to\nu\nu'S_k)=\Gamma_{S_l\to\nu\nu'S_k}/\Gamma_{S_l}$\,
from the rates derived in Appendix\,\,\ref{Sdecays}.

The LEP\,\,II experiments on \,$e^+e^-\to\gamma\,\slash\!\!\!\!E$\, had c.m. energies within
the range \,130-207\,\,GeV,\, and the observed cross-sections vary also with cuts
on the photon energy $E_\gamma$ and angle $\theta_\gamma$ relative to the beam direction.
From a collection of these data~\cite{lep:ee2gnn} tabulated in~Ref.\,\cite{Chiang:2012ww},
one can see that the majority of the measured and SM values of the cross section agree
with each other at the one-sigma level.
Consequently, to bound the \,$eeSS'$\, couplings, we may require
$\sigma_{e\bar e\to\gamma S\bar S'\to\gamma\,\slash\!\!\!\!E}$ not to exceed the corresponding
one-sigma empirical errors (after combining the statistical and systematic errors in quadrature).

Applying this condition and assuming as before that only one of the $C$ couplings in
Eq.\,(\ref{CO}) is nonzero at a time, for the coupling choices in Eq.\,(\ref{ccc}) we find
that $\kappa_{LR}^{}/\Lambda^2$ does not get any meaningful limitations from the LEP\,\,II
measurements, which is not unexpected because the resulting $eeSS'$ interaction is
suppressed by the electron mass, as Eq.\,(\ref{ccc[ee2ss]}) indicates.
On the other hand, they do translate into moderate restraints on $\kappa_{L(R)}/\Lambda^2$.
More precisely, from the data, we infer the dotted curves shown in Figure\,\,\ref{Lambda[L,R]}
which represent lower limits on $\tilde\Lambda$ and therefore reduce the parameter
space consistent with the observed relic abundance (the solid thin bands), so that now
\,$m_{S_3}\mbox{\footnotesize\;$\lesssim$\;}24\,(43)$\,\,GeV\, is excluded for
\,$\epsilon=L\,(R)$.\,
It is clear from this simple exercise that future $e^+e^-$ machines with greater energies and
luminosities, such as the International Linear Collider\,\,\cite{ilc}, can be expected to
probe more stringently this new-physics scenario, if they detect no signals beyond the SM.

\begin{figure}[b]
\includegraphics[width=93mm]{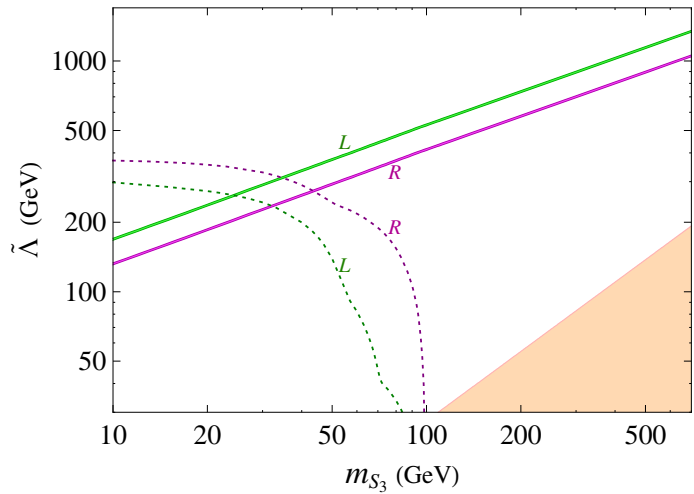}\vspace{-7pt}
\caption{Values of \,$\tilde\Lambda=\Lambda/|\kappa_\epsilon|^{1/2}$\, for
\,$\epsilon=L$ and $R$\, which are compatible with the observed relic abundance
(solid thin bands), compared to lower limits on $\tilde\Lambda$ inferred from
measurements of \,$e^+e^-\to\gamma\,\slash\!\!\!\!E$\, at LEP\,\,II
(dotted curves).\label{Lambda[L,R]}}
\end{figure}

Another important implication of the dimension-6 effective \,$\ell\ell'S S'$\,
interactions is that they can give rise to the flavor-changing decay
\,$\ell_a^-\to\ell_b^-\ell_c^-\ell_d^+$\, via one-loop diagrams
involving internal $S_{k,l}$ if at least one of the couplings is flavor violating.
Such decays have been searched for over the years, but with null results so far,
leading to increasingly severe bounds on their branching ratios \cite{pdg}.
Consequently, their data may give rise to substantial restrictions on the couplings.

We again assume that only one of the couplings in Eq.\,(\ref{CO}) is contributing at a time.
Since $C^{\scriptscriptstyle R}$ conserves flavor, only $C^{\scriptscriptstyle LR}$ and
$C^{\scriptscriptstyle L}$ as specified in Eq.\,(\ref{ccc}) are pertinent.
Thus we can express the amplitudes for \,$\ell_a^-\to\ell_b^-\ell_c^-\ell_d^+$\,
in each case as
\begin{eqnarray} \label{MLR}
{\cal M}_{\ell_a^{}\to\ell_b^{}\ell_c^{}\bar\ell_d^{}}^{\scriptscriptstyle LR} &=\,&
\frac{{\cal I}\bigl(m_{S_k},m_{S_l}\bigr)v^2}{16_{\;}\!\pi_{\,\,}^2\Lambda^4}\,
{\cal U}_{pk\,}^*{\cal U}_{sk\,}^{}{\cal U}_{ql\,}^{}{\cal U}_{rl}^* \bigl[ \bar u_c^{}
\bigl(C_{capq}^{\scriptscriptstyle LR}P_R^{}+C_{acqp}^{{\scriptscriptstyle LR}*}P_L^{}\bigr)
u_{a\,}^{} \bar u_b^{}
\bigl(C_{bdrs}^{\scriptscriptstyle LR}P_R^{}+C_{dbsr}^{{\scriptscriptstyle LR}*}P_L^{}\bigr)
v_d^{}
\nonumber \\ && \hspace{30ex} -\; (b\leftrightarrow c) \bigr] \,,
\\ \label{ML}
{\cal M}_{\ell_a^{}\to\ell_b^{}\ell_c^{}\bar\ell_d^{}}^{\scriptscriptstyle L} &=\,&
\frac{{\cal J}\bigl(m_{S_k},m_{S_l}\bigr)}
{8_{\;}\!\pi_{\,\,}^2\Lambda^4}\, {\cal U}_{pk\,}^*{\cal U}_{rk}^{}\,{\cal U}_{ql\,}^{}
{\cal U}_{sl}^*\, \bigl( C_{capq\,}^{\scriptscriptstyle L}C_{bdsr}^{\scriptscriptstyle L}
+ C_{bapq\,}^{\scriptscriptstyle L} C_{cdsr}^{\scriptscriptstyle L} \bigr)
\bar u_c^{}\gamma^\rho P_L^{}u_a^{}\,\bar u_b^{}\gamma_\rho^{}P_L^{}v_d^{} \;,
\end{eqnarray}
where \,$k,l,p,q,r,s=1,2,3$\, are summed over and $\cal I$ and $\cal J$ are loop functions.
With the choices of $C^{\scriptscriptstyle LR,L}$ in Eq.\,(\ref{ccc}), we arrive at
\begin{eqnarray}
{\cal I}(m,n) \,\,=\,\, \frac{m^2\ln(n/m)}{m^2-n^2} ~, ~~~~~~~
{\cal J}(m,n) \,\,=\,\, m^2 \biggl(\ln\frac{\Lambda}{n}+\frac{1}{4}\biggr)
\,+\, \frac{m^4\ln(n/m)}{m^2-n^2} ~,
\end{eqnarray}
where we have dropped terms that vanished after $k$ is summed over in Eqs.\,\,(\ref{MLR}) and
(\ref{ML}) due to \,$a\neq b,c,d$\, and the unitarity of $\,\cal U$.
We have also taken the cutoff in the loop integration to be the same as the scale $\Lambda$
and neglected the momenta of the external particles.

Upon comparing the resulting branching ratio of \,$\ell_a^-\to\ell_b^-\ell_c^-\ell_d^+$\, to
its measured bound, one can then derive a limit on \,$\Lambda/|\kappa_{LR}|^{1/2}$,\,
assuming that only $C^{\scriptscriptstyle LR}$ is nonzero.
The relevant modes are \,$\mu^-\to e^-e^-e^+$\, and
\,$\tau^-\to e^-e^-e^+,\mu^-\mu^-\mu^+,e^-e^-\mu^+,\mu^-\mu^-e^+,\mu^-e^-e^+,e^-\mu^-\mu^+$,\,
for which only experimental bounds on the branching ratios are available.
Although the strictest among them is
\,${\cal B}\bigl(\mu^-\to e^-e^-e^+\bigr){}_{\rm exp}^{}<1.0\times10^{-12}$,\, we find that
\,${\cal B}\bigl(\tau^-\to\mu^-\mu^-\mu^+\bigr){}_{\rm exp}^{}<2.1\times10^{-8}$\,~\cite{pdg}
yields the strongest constraint, namely
\begin{eqnarray}
\frac{\Lambda}{|\kappa_{LR}^{}|^{1/2}} \,\,>\,\, 11\rm~GeV \,,
\end{eqnarray}
which is consistent with the $LR$ curve in Figure\,\,\ref{tLambda}.
This is mainly due to the enhancement from the lepton mass factor in the rate of
\,$\tau^-\to\mu^-\mu^-\mu^+$,\, as can be seen from the expressions for
\,$\ell^{\prime-}\to\ell^-\ell^-\ell^+$\, rates collected in Appendix\,\,\ref{l->lll}.

\begin{figure}[t]
\includegraphics[width=93mm]{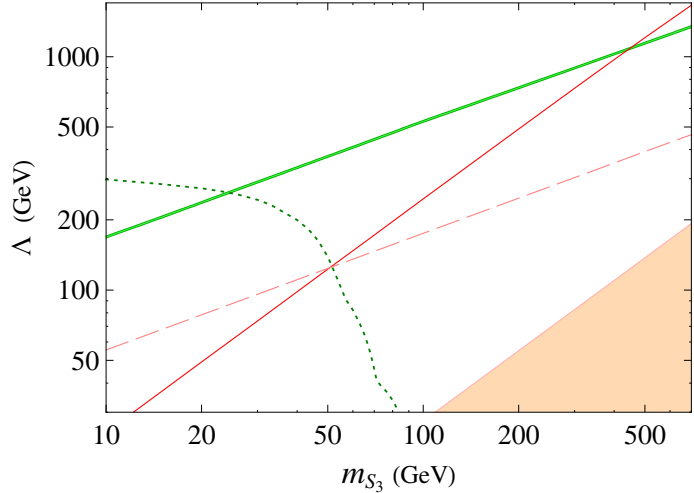}\vspace{-7pt}
\caption{Values of $\Lambda$ for $C^{\scriptscriptstyle L}$ in Eq.\,(\ref{ccc})
with \,$\kappa_L^{}=1$\, which fulfill the relic density constraint
(green thin band), compared to the lower limits on $\Lambda$ from the LEP\,\,II data
on \,$e^+e^-\to\gamma\,\slash\!\!\!\!E$\, (green dotted curve) and from searches for
\,$\mu^-\to e^-e^-e^+$\, (red solid curve) and \,$\tau^-\to\mu^-\mu^-e^+$\,
(red dashed curve).\label{LambdaL}}
\end{figure}

If instead only $C^{\scriptscriptstyle L}$ in Eq.\,(\ref{ccc}) is present,
${\cal B}\bigl(\mu^-\to e^-e^-e^+\bigr){}_{\rm exp}^{}$ turns out to impose the most stringent
constraint among these decays in the \,$m_{S_3}\mbox{\footnotesize\;$\gtrsim$\;}50$\,GeV\,
region, whereas for lower masses
\,${\cal B}\bigl(\tau^-\to\mu^-\mu^-e^+\bigr){}_{\rm exp}^{}<1.7\times10^{-8}$\,~\cite{pdg}
is the most restrictive.
The formulas for their rates are also listed in Appendix\,\,\ref{l->lll}.
In Figure$\;$\ref{LambdaL} we depict the resulting lower-limits on $\Lambda$.
In this case, we set \,$\kappa_L^{}=1$\, due to the ${\rm ln}\Lambda$ dependence of
the \,$\mu^-\to e^-e^-e^+$\, rate.
The plot reveals that above \,$m_{S_3}\sim500$\,GeV\, the $\Lambda$ values consistent with
the observed relic density are in conflict with the bound from the \,$\mu^-\to e^-e^-e^+$\, data.
This significantly shrinks the allowed parameter space already decreased by the restraint from
the LEP\,\,II measurements.

\subsection{Flavor-violating Higgs decay\label{h2ll'}}

The recently discovered Higgs boson can potentially offer a window into physics beyond
the~SM.
The presence of new particles can bring about modifications to the standard decay modes
of the Higgs and/or cause it to undergo exotic decays~\cite{Curtin:2013fra}.
As data from the LHC continues to accumulate with increasing precision, they may
uncover clues of new physics in the Higgs couplings.

The CMS Collaboration~\cite{cms:h2mt} has recently reported the detection of a slight
excess of \,$h\to\mu^\pm\tau^\mp$\, events with a significance of 2.5$\sigma$.
If interpreted as a signal, the result corresponds to a branching fraction of
\,${\cal B}(h\to\mu\tau)={\cal B}(h\to\mu^-\tau^+)+{\cal B}(h\to\mu^+\tau^-)=
\bigl(0.89^{+0.40}_{-0.37}\bigr)$\%,\,
but as a statistical fluctuation it translates into a limit of
\,${\cal B}(h\to\mu\tau)<1.57$\%\, at 95\% CL~\cite{cms:h2mt}.
It is too early to draw a definite conclusion from this finding, but it would constitute
clear evidence of physics beyond the SM if substantiated by future measurements.
Assuming that the tentative signal hint is true, we investigate whether the $S_k$
interactions could effect such an exotic Higgs decay within the allowed parameter
space.\footnote{The CMS excess has also been addressed in the contexts of other
new-physics scenarios~\cite{h2mt}.}

One can write the amplitude for \,$h\to\ell_b^-\ell_d^+$\, as
\begin{eqnarray}
{\cal M}_{h\to\ell_b\bar\ell_d} \,\,=\,\,
\frac{y_{bd}^{\scriptscriptstyle\rm SM}+y_{bd}^{\rm new}}{v}\,
\bar u_b^{} \bigl(m_{\ell_b}P_L^{}+m_{\ell_d}P_R^{}\bigr)v_d^{} \;,
\end{eqnarray}
corresponding to the rate
\begin{eqnarray}
\Gamma_{h\to\ell_b\bar\ell_d} \,\,=\,\, \frac{m_{h\,}^{}
\bigl|y_{bd}^{\scriptscriptstyle\rm SM}+y_{bd}^{\rm new}\bigr|\raisebox{1pt}{$^2$}}
{16\pi v^2} \bigl(m_{\ell_b}^2+m_{\ell_d}^2\bigr) \;,
\end{eqnarray}
where \,$y_{bd}^{\scriptscriptstyle\rm SM}=\delta_{bd}^{}$\, in the SM at tree level
and $y_{bd}^{\rm new}$ stands for the new contribution.
The main contribution to $y_{bd}^{\rm new}$ originates from a one-loop diagram involving
internal $S_k$, one $hS_k S_k$ vertex from the renormalizable Lagrangian in~Eq.\,(\ref{hss}),
and one $S_k S_k\ell\ell'$ vertex from a dimension-six operator in~Eq.\,(\ref{L'}).
It turns out that for the latter coupling only $O^{\scriptscriptstyle LR}$ matters, as the loop
contributions of $O^{\scriptscriptstyle L,R}$ vanish.
Thus, focusing on the case in which \,$m_h^{}<2m_{S_k}$\, and $C^{\scriptscriptstyle LR}$
is given by Eq.\,(\ref{ccc}), we obtain
\begin{eqnarray} \label{ybd}
y_{bd}^{\rm new} \,\,=\,\,
\frac{\kappa_{LR\,}^{}\lambda_k^{}\,{\cal U}_{dk\,}^*{\cal U}_{bk\,}^{}v^2}
{16\pi_{\,\,}^2\Lambda^2} \left( \ln\frac{\Lambda^2}{m_{S_k}^2} + 1
- 2\sqrt{\frac{4 m_{S_k}^2}{m_h^2}-1}\; \sin^{-1}\frac{m_h^{}}{2m_{S_k}} \right) ,
\end{eqnarray}
where summation over \,$k=1,2,3$\, is implicit and we have again taken the cutoff in
the loop integration to be the same as $\Lambda$.
The $S_k S_k\ell\ell'$ coupling alone can generate one-loop contributions to the off-diagonal
elements of the charged-lepton mass matrix, but we estimate their impact on its eigenvalues
to be small.
Therefore, $y_{bd}^{\rm new}$ in Eq.\,(\ref{ybd}) is largely unaffected as the leading
contribution of $S_k$ to~\,$h\to\ell_b^-\ell_d^+$.\,

Since \,$h\to\tau^+\tau^-,\mu^+\mu^-$\, also receive the $S_k$ contributions
in Eq.\,(\ref{ybd}), we need to take into account the relevant data.
The ATLAS and CMS Collaborations have reported the observations of \,$h\to\tau^+\tau^-$\,
and measured its signal strength to be
\,$\sigma/\sigma_{\scriptscriptstyle\rm SM}^{}=1.42^{+0.44}_{-0.38}$
and $0.91\pm0.27$,\, respectively~\cite{CMS:2014ega,atlas:h2tt}.
In contrast, the only experimental information on \,$h\to\mu^+\mu^-$\, are the bounds
\,${\cal B}(h\to\mu\bar\mu)<1.5\times10^{-3}$ and $1.6\times10^{-3}$\, from ATLAS and
CMS, respectively~\cite{atlas:h2mm,cms:h2mm}.
In view of these data, we demand the $S_k$ contributions to respect
\begin{eqnarray} \label{h2lldata}
0.7 \,\,<\,\,
\frac{\Gamma_{h\to\tau\bar\tau}}{\Gamma_{h\to\tau\bar\tau}^{\scriptscriptstyle\rm SM}}
\,\,<\,\, 1.8 ~, ~~~~~~~
\frac{\Gamma_{h\to\mu\bar\mu}}{\Gamma_{h\to\mu\bar\mu}^{\scriptscriptstyle\rm SM}}
\,\,<\,\, 6.7 ~,
\end{eqnarray}
where \,$\Gamma_{h\to\tau\bar\tau}^{\scriptscriptstyle\rm SM}=257$\,keV\,
and \,$\Gamma_{h\to\mu\bar\mu}^{\scriptscriptstyle\rm SM}=894$\,eV\, \cite{lhctwiki}
for \,$m_h^{}=125.1$\,GeV.\,

Due to the ln$\Lambda$ dependence of $y_{bd}^{\rm new}$, we also set \,$\kappa_{LR}^{}=1$.\,
It follows that, for illustration, we can select \,$\bigl(m_{S_3},\Lambda\bigr)=(70,79)$
and (200,78)\,\,GeV\, from the $LR$ (blue) curve in Figure$\;$\ref{tLambda},
implying that we have assumed $\lambda_3^{}$ to be negligible.
Choosing also \,$\lambda_1=\lambda_2$\, for simplification, we find that for
\,$\bigl(m_{S_3},\Lambda\bigr)=(70,79)$\,GeV\, the $S_k$ contributions lead to
\,$0.0026>\bigl|y_{\mu\tau,\tau\mu}^{\rm new}\bigr|m_\tau^{}/v>0.0021$,\,
or~\,$0.79\%>{\cal B}(h\to\mu\tau)>0.52\%$,\,
compatible with the range of the CMS finding on
the potential signal~\cite{cms:h2mt}, if \,$-7.2<\lambda_1<-5.8$.\,
For \,$\bigl(m_{S_3},\Lambda\bigr)=(200,78)$\,GeV,\, we obtain the same ${\cal B}(h\to\mu\tau)$
range if \,$-2.9<\lambda_1<-2.4$.\,
All these numbers correspond to
\,$1.6<\Gamma_{h\to\tau\bar\tau}/\Gamma_{h\to\tau\bar\tau}^{\scriptscriptstyle\rm SM}<1.8$\,
and
\,$1.8<\Gamma_{h\to\mu\bar\mu}/\Gamma_{h\to\mu\bar\mu}^{\scriptscriptstyle\rm SM}<2.0$,\,
which conform to the conditions in Eq.\,(\ref{h2lldata}) and are therefore testable soon with
forthcoming data from the LHC.
Moreover, we determine that \,$\Gamma_{h\to e\tau}=0.053_{\,}\Gamma_{h\to\mu\tau}$.\,
Although the preferred values of \,$|\lambda_{1,2}|$\, seem to be sizable, they are still
below the perturbativity limit of $4\pi$ mentioned earlier.
It is worth noting that the \,$\bigl|y_{\mu\tau,\tau\mu}^{\rm new}\bigr|m_\tau^{}/v$\,
numbers above are below the upper limit of 0.016 inferred from the measured bound on
the \,$\tau\to\mu\gamma$\, decay~\cite{h2ll'}.

We have seen from the limited exercises performed in this paper that the MFV framework offers
a systematic way to explore potential relations between DM, neutrinos, and the Higgs boson through
a variety of processes which can be checked experimentally.
More sophisticated choices of the coefficients $C^{\scriptscriptstyle L,R,LR}$ than those in
Eq.\,(\ref{ccc}) would then allow the examination of a~greater number of leptonic observables.

\section{Conclusions\label{conclusion}}

We have considered DM which is a singlet under the SM gauge group and a member of
a~scalar triplet under the lepton flavor group.
The triplet is odd under an extra $Z_2$ symmetry which renders the DM candidate stable.
We apply the MFV principle to all the lepton-flavored particles in the theory which
includes three right-handed neutrinos taking part in the seesaw mechanism for neutrino
mass generation.
The new scalars couple to SM particles via Higgs-portal renormalizable interactions
and dimension-six operators involving leptons.
The MFV framework allows us to make interesting phenomenological connections between
the DM, Higgs, and lepton sectors.
We examine restrictions on the new scalars from the Higgs boson data, observed relic
density, DM\,\,direct searches, LEP\,\,II measurements on $e^+e^-$ scattering into
a photon plus missing energy, and experimental bounds on flavor-violating lepton decays.
We obtain viable parameter space that can be probed further by future experiments.
Our simple choices of the new scalars' effective couplings illustrate how various data
can constrain them in complementary ways.
We also explore whether the scalar interactions can account for the tentative hint
of the Higgs flavor-violating decay \,$h\to\mu\tau$\, recently detected in the CMS experiment.
Their contributions, occurring at the one-loop level, can give rise to a decay rate compatible
with the CMS finding and at the same time fulfill requirements from other Higgs data.
If it is not confirmed by upcoming measurements, the acquired data will place stronger
limitations on the considered scenario of lepton-flavored DM with MFV.
Last but not least, it is clear from our analysis that next-generation $e^+e^-$ machines
with high energies and luminosities, such as the International Linear Collider, have
considerable potential for testing different aspects of this kind of new physics in
greater detail.

\acknowledgments

We would like to thank Xiao-Gang He for helpful comments.
We also thank Brian Batell for remarks concerning the stability of lepton-flavored dark matter.
This work was supported in part by the MOE Academic Excellence Program (Grant No. 102R891505)
and the NCTS.

\appendix

\section{Cross section of $\bm{e^+e^-\to\gamma S_k^{}S_l^*}$\label{csr}}

For the scattering \,$e^-(p)\,e^+(\bar p)\to\gamma(\textsc{k})\,S_k^{}(q)\,S_l^*(\bar q)$,\,
we define the Lorentz-invariant kinematical variables
\begin{eqnarray} \label{stuk}
s &\,=\,& (p+\bar p)^2 \,, \hspace{7ex} s' \,\,=\,\, (q+\bar q)^2 \,, \hspace{7ex}
t \,\,=\,\, (\bar p-\bar q)^2 \,, \hspace{7ex}\! t' \,\,=\,\, (p-q)^2 \,, \nonumber \\
u &\,=\,& (\bar p-q)^2 \,, \hspace{7ex} u' \,\,=\,\, (p-\bar q)^2 \,, \hspace{6ex}
w \,\,=\,\, 2_{\,}\textsc{k}\cdot p \,, \hspace{8ex}
\bar w \,\,=\,\, 2_{\,}\textsc{k}\cdot\bar p \,. ~~~
\end{eqnarray}
We derive its amplitude ${\cal M}_{e\bar e\to\gamma S_k\bar S_l}$ from two diagrams
each with an \,$e^-e^+\to S_k^{}S_l^*$\, vertex and the photon radiated from
the $e^-$ or $e^+$ leg.
Thus, in the limit of massless $e^\pm$,
\begin{eqnarray} \label{ee2gSS'}
{\cal M}_{e\bar e\to\gamma S_k\bar S_l}^{} &=&
\frac{\sqrt{4\alpha_{\;\!}\pi}}{\Lambda^2}\,\bar v_{\bar e}^{} \bigg[ (\slash{\!\!\!\bar q}-\slash{\!\!\!q})
\bigl(\textsc{c}_{kl}^{\scriptscriptstyle L}P_L^{}
+ \textsc{c}_{kl}^{\scriptscriptstyle R}P_R^{}\bigr)
- \frac{v}{\sqrt2} \bigl(\textsc{c}_{kl}^{\scriptscriptstyle LR}P_R^{} +
\bar{\textsc{c}}_{kl}^{\scriptscriptstyle LR}P_L^{}\bigr) \bigg]
\frac{\slash{\!\!\!p}-\slash{\!\!\!\textsc{k}}}{w}\,\slash{\!\!\!\varepsilon}^*u_e^{}
\nonumber \\ && \!\!\! -\;
\frac{\sqrt{4\alpha_{\;\!}\pi}}{\Lambda^2}\,\bar v_{\bar e}^{}\,\slash{\!\!\!\varepsilon}^*\,
\frac{\slash{\!\!\!\bar p}-\slash{\!\!\!\textsc{k}}}{\bar w}
\bigg[ (\slash{\!\!\!\bar q}-\slash{\!\!\!q}) \bigl(\textsc{c}_{kl}^{\scriptscriptstyle L}P_L^{}
+ \textsc{c}_{kl}^{\scriptscriptstyle R}P_R^{}\bigr)
- \frac{v}{\sqrt2} \bigl( \textsc{c}_{kl}^{\scriptscriptstyle LR}P_R^{} +
\bar{\textsc{c}}_{kl}^{\scriptscriptstyle LR}P_L^{}\bigr) \bigg] u_e^{} \;, ~~~~~~~
\end{eqnarray}
where \,$\alpha=1/128$\, is the fine-structure constant,
\begin{eqnarray}
\textsc{c}_{kl}^\epsilon \,\,=\,\, \raisebox{3pt}{\footnotesize$\displaystyle\sum_{n,o}$}\;
{\cal U}_{kn}^\dagger\,{\cal U}_{ol}^{}\,C_{11no}^\epsilon \;, ~~~~~
\epsilon \,\,=\,\, L,R,LR \;, ~~~~~~~
\bar{\textsc{c}}_{kl}^{\scriptscriptstyle LR} \,\,=\,\,
\bigl(\textsc{c}_{lk}^{\scriptscriptstyle LR}\bigr)^* \,.
\end{eqnarray}
Hence for the choices in Eq.\,(\ref{ccc})
\begin{eqnarray} \label{ccc[ee2ss]}
\textsc{c}_{kl}^{\scriptscriptstyle L} \,\,=\,\,
2 \kappa_L^{}\,{\cal U}_{1k\,}^*{\cal U}_{1l}^{} \;, ~~~~~~~
\textsc{c}_{kl}^{\scriptscriptstyle R} \,\,=\,\, \kappa_{R\,}^{}\delta_{kl}^{} \;, ~~~~~~~
\textsc{c}_{kl}^{\scriptscriptstyle LR} \,\,=\,\, \frac{\sqrt2\,\kappa_{LR\,}^{}m_e^{}}{v}\;
{\cal U}_{1k\,}^*{\cal U}_{1l}^{} \;.
\end{eqnarray}
It is easy to check that \,${\cal M}_{e\bar e\to\gamma S_k\bar S_l}$\, respects
electromagnetic gauge invariance.
Averaging (summing) its absolute square over the initial (final) spins, one then obtains
\begin{eqnarray} \label{M2[ee2gss]}
\overline{\big|{\cal M}_{e\bar e\to\gamma S_k\bar S_l}\big|\raisebox{1pt}{$^2$}}  &=&
\frac{2\alpha_{\;\!}\pi_{\;\!}\big(|\textsc{c}_{kl}^{\scriptscriptstyle L}|^2
+ |\textsc{c}_{kl}^{\scriptscriptstyle R}|^2\big)}
{\Lambda^{4\,}w_{\;\!}\bar w} \Big\{ 2 \Big(m_{S_k^{}}^2-m_{S_l^{}}^2\Big)
\Big[ m_{S_k^{}}^2 s-m_{S_l^{}}^2 s+(t-u)w-\big(t'-u'\big)\bar w \Big]
\nonumber \\ && \hspace{5em}
+\, \bigl(w^2+\bar w^2+2s s'\big)\Big(s'-2m_{S_k^{}}^2-2m_{S_l^{}}^2\Big)
- s'(t-u)^2-s'\big(t'-u'\big)\raisebox{1pt}{$^2$} \Big\}
\nonumber \\ && +\;
\frac{\alpha_{\;\!}\pi_{\;\!}v^2}{\Lambda^{4\,}w_{\;\!}\bar w} \bigl(
|\textsc{c}_{kl}^{\scriptscriptstyle LR}|^2 + |\textsc{c}_{lk}^{\scriptscriptstyle LR}|^2
\bigr) \bigl(s^2+s^{\prime2}\bigr) \,.
\end{eqnarray}
This leads to the cross section
\begin{eqnarray} \label{exact}
\sigma_{e\bar e\to\gamma S_k\bar S_l}^{} = \int
\frac{E_\gamma\,dE_\gamma\,d(\cos\theta_\gamma)\,d\bar\Omega_S}{2 (4\pi)^4\,s}
\sqrt{1-\frac{2m_{S_k}^2+2m_{S_l}^2}{s-2E_\gamma\sqrt s}
+ \Biggl(\frac{m_{S_k}^2-m_{S_l}^2}{s-2E_\gamma\sqrt s}\Biggr)\raisebox{12pt}{$^{\!\!\!2}$}} ~
\overline{|{\cal M}_{e\bar e\to\gamma S_k\bar S_l}|^2} \;, ~~~~
\end{eqnarray}
where $E_\gamma^{}$ and $\theta_\gamma^{}$ are the photon energy and angle with respect to
the $e^-$ or $e^+$ beam direction in the c.m. frame of the $e^+e^-$ pair,
$\bar\Omega_S^{}$ denotes the solid angle of either $S_k^{}$ or $S_l^*$ in the c.m. frame of
the $S_k^{}S_l^*$ pair.
The photon energy range is
\begin{eqnarray} \label{Egrange}
E_\gamma^{\rm min} \,\,\le\,\, E_\gamma^{}\,\,\le\,\, E_\gamma^{\rm max}
\,\,=\,\, \frac{s-(m_{S_k}+m_{S_l})^2}{2\sqrt s} \;,
\end{eqnarray}
where $E_\gamma^{\rm min}$ is an experimental cut.
In the numerical evaluation of the integral, the $\theta_\gamma^{}$ range is also subject to cuts.

It is worth mentioning that one could alternatively estimate
\,$\sigma_{e\bar e\to\gamma S_k\bar S_l}$\, in the so-called
radiator approximation~\cite{Nicrosini:1988hw}.
It is given by
\begin{eqnarray} \label{approx} & \displaystyle
\sigma_{e\bar e\to\gamma S_k\bar S_l}^{} \,\,=\,\, \int dc_\gamma^{}\,dx_\gamma^{}\,
{\cal H}\big(c_\gamma^{},x_\gamma^{};s\big)\, \hat\sigma(\hat s) \,,
& \\ & \displaystyle
c_\gamma^{} \,=\, \cos\theta_\gamma^{} \,, \hspace{4ex}
x_\gamma^{} \,=\, \frac{2 E_\gamma}{\sqrt s} \,, \hspace{4ex}
{\cal H}\big(c_\gamma^{},x_\gamma^{};s\big) \,=\, \frac{\alpha}{\pi}\;
\frac{\big(2-x_\gamma^{}\big)^2+c_\gamma^2 x_\gamma^2}
{2\big(1-c_\gamma^2\big)\,x_\gamma^{}} \,, \hspace{4ex}
\hat s \,=\, s-s x_\gamma^{} \,, & ~~~~ \nonumber
\end{eqnarray}
where $\hat\sigma(\hat s)$ stands for the cross section of the simpler reaction
\,$e^+e^-\to S_k^{}S_l^*$,\,
\begin{eqnarray}
\hat\sigma(\hat s) \,\,=\,\, \frac{{\cal K}^{\frac{3}{2}}\bigl(\hat s,m_{S_k}^2,m_{S_l}^2\bigr)}
{96\Lambda^{4\,}\pi_{\,}\hat s^2} \bigl(|\textsc{c}_{kl}^{\scriptscriptstyle L}|^2
+ |\textsc{c}_{kl}^{\scriptscriptstyle R}|^2\bigr)
\,+\, \frac{{\cal K}^{\frac{1}{2}}\bigl(\hat s,m_{S_k}^2,m_{S_l}^2\bigr)_{\,}v^2}
{128\Lambda^{4\,}\pi_{\,}\hat s} \bigl(|\textsc{c}_{kl}^{\scriptscriptstyle LR}|^2
+ |\textsc{c}_{lk}^{\scriptscriptstyle LR}|^2\bigr) \;.
\end{eqnarray}
With this method, the $\textsc{c}_{kl}^{{\scriptscriptstyle L},\scriptscriptstyle R}$
contributions to $\sigma_{e\bar e\to\gamma S_k\bar S_l}$ turn out to be exactly the same as
their counterparts in Eq.\,(\ref{exact}), whereas the $\textsc{c}_{kl}^{\scriptscriptstyle LR}$
terms would yield numbers lower by no more than several percent.

\section{Decays of $\bm{S}$ particles\label{Sdecays}}

The decay of $S_l$ into $S_k$ plus charged leptons if kinematically permitted may arise from
the operators $O^{{\scriptscriptstyle L},{\scriptscriptstyle R},{\scriptscriptstyle LR}}$ in
Eq.\,(\ref{L'}), depending on the specifics of the couplings.
For $O^{\scriptscriptstyle L}$, the final leptons can also be neutrinos instead.
The amplitudes for \,$S_l^{}(q)\to S_k^{}(p)\,\ell_b^-\ell_d^+$\, and
\,$S_l^{}(q)\to S_k^{}(p)\,\nu_b^{}\nu_d^{}$\, are then
\begin{eqnarray} \label{S2llS}
{\cal M}_{S_l^{}\to\ell_b^{}\bar\ell_d^{}S_k^{}}^{} &\,=\,& \frac{-1}{\Lambda^2}\,\bar u_{b\,}^{}
\Bigg[ \frac{v}{\sqrt2} \big( {\sf c}_{bdkl}^{\scriptscriptstyle LR}P_R^{} +
\bar{\sf c}_{bdkl}^{\scriptscriptstyle LR}P_L^{}\big)
\,+\, (p+q)_\rho^{}\,\gamma^\rho \big({\sf c}_{bdkl}^{\scriptscriptstyle L}P_L^{}
+ {\sf c}_{bdkl}^{\scriptscriptstyle R}P_R^{}\big) \Bigg] v_d^{} \;, ~~~~ ~~~
\\ \label{S2nnS}
{\cal M}_{S_l^{}\to\nu_b^{}\nu_d^{}S_k^{}}^{} &\,=\,& \frac{-1}{\Lambda^2} (p+q)_\rho^{}\,
\bar u_{b\,}^{}\gamma^\rho\big({\sf c}_{bdkl}^{\scriptscriptstyle L}P_L^{}
- {\sf c}_{dbkl}^{\scriptscriptstyle L}P_R^{}\big)v_d^{} \;,
\end{eqnarray}
where
\begin{eqnarray}
{\sf c}_{bdkl}^\epsilon \,\,=\,\, \raisebox{3pt}{\footnotesize$\displaystyle\sum_{n,o}$}
\; {\cal U}_{kn}^\dagger\,{\cal U}_{ol}^{}\, C_{bdno}^\epsilon \;, ~~~~~~~
\bar{\sf c}_{bdkl}^{\scriptscriptstyle LR} \,\,=\,\,
\bigl({\sf c}_{dblk}^{\scriptscriptstyle LR}\bigr)^* \,.
\end{eqnarray}
Thus for the choices in Eq.\,(\ref{ccc})
\begin{eqnarray} \label{ccc[s2lls]}
\textsf{c}_{bdkl}^{\scriptscriptstyle L} \,\,=\,\,
2 \kappa_L^{}\,{\cal U}_{dk\,}^*{\cal U}_{bl}^{} \;, ~~~~~~~
\textsf{c}_{bdkl}^{\scriptscriptstyle R} \,\,=\,\, \kappa_{R\,}^{}\delta_{bd}^{}\delta_{kl}^{} \;, ~~~~~~~
\textsf{c}_{bdkl}^{\scriptscriptstyle LR} \,\,=\,\, \frac{\sqrt2\,\kappa_{LR\,}^{}m_{\ell_d}^{}}{v}\;
{\cal U}_{dk\,}^*{\cal U}_{bl}^{} \;,
\end{eqnarray}
and so with the above $\textsf{c}_{bdkl}^{\scriptscriptstyle R}$ alone $S_{1,2,3}$ are all stable.
From Eqs.\,\,(\ref{S2llS}) and (\ref{S2nnS}), the decay rates for negligible lepton masses are
\begin{eqnarray}
\Gamma_{S_l^{}\to\ell_b^{}\bar\ell_d^{}S_k^{}}^{} &=&
\frac{\bigl(|{\sf c}_{bdkl}^{\scriptscriptstyle LR}|^2
+ |{\sf c}_{dblk}^{\scriptscriptstyle LR}|^2\bigr)v^2}
{3072_{\,}\Lambda^4\pi^{3\,}m_{S_l}^3} \Biggl[
\bigl(m_{S_l}^2-m_{S_k}^2\bigr)\bigl(m_{S_l}^4+10_{\,}m_{S_k}^2m_{S_l}^2+m_{S_k}^4\bigr)
\nonumber \\ && \hspace{22ex} -\;
12_{\,}m_{S_k}^2 m_{S_l}^2\bigl(m_{S_k}^2+m_{S_l}^2\bigr) \ln\frac{m_{S_l}^{}}{m_{S_k}^{}} \Biggr]
\nonumber \\ && \!\!\! +\;
\frac{|{\sf c}_{bdkl}^{\scriptscriptstyle L}|^2+|{\sf c}_{bdkl}^{\scriptscriptstyle R}|^2}
{1536_{\,}\Lambda^4\pi^{3\,}m_{S_l}^3} \Biggl[
\bigl(m_{S_l}^4-m_{S_k}^4\bigr)\bigl(m_{S_l}^4-8_{\,}m_{S_k}^2m_{S_l}^2+m_{S_k}^4\bigr)
+ 24_{\,}m_{S_k}^4 m_{S_l}^4 \ln\frac{m_{S_l}^{}}{m_{S_k}^{}} \Biggr] \,,
\nonumber \\ \\
\Gamma_{S_l^{}\to\nu\nu'S_k^{}}^{} &=& \mbox{$\frac{1}{2}$}\;
\raisebox{3pt}{\footnotesize$\displaystyle\sum_{b,d}$}\;\Gamma_{S_l^{}\to\nu_b^{}\nu_d^{}S_k^{}}^{}
\nonumber \\ && \hspace{-4ex} =\,
\frac{\sum_{b,d}^{}|{\sf c}_{bdkl}^{\scriptscriptstyle L}|^2}{1536_{\,}\Lambda^4\pi^{3\,}m_{S_l}^3}
\Biggl[ \bigl(m_{S_l}^4-m_{S_k}^4\bigr)\bigl(m_{S_l}^4-8_{\,}m_{S_k}^2m_{S_l}^2+m_{S_k}^4\bigr)
+ 24_{\,}m_{S_k}^4 m_{S_l}^4 \ln\frac{m_{S_l}^{}}{m_{S_k}^{}} \Biggr] \,,
\end{eqnarray}
where the factor of $\frac{1}{2}$ in $\Gamma_{S_l\to\nu\nu'S_k}$ accounts for the identical
Majorana neutrinos in the final states of channels with \,$b=d$\, and prevents double counting
of contributions with \,$b\neq d$.\,
In the numerical evaluation of $\Gamma_{S_l\to\ell_b\bar\ell_d S_k}$, we do not neglect
the lepton masses.
For our $m_{S_k}$ choices, these three-body modes dominate the total widths of $S_{1,2}$,
and so we can approximate them to be
\,$\Gamma_{S_1}=\Gamma_{S_1\to\nu\nu'S_3}+\Gamma_{S_1\to\ell\bar\ell'S_3}$\, and
\,$\Gamma_{S_2}=\Gamma_{S_2\to\nu\nu'S_1}+\Gamma_{S_2\to\nu\nu'S_3}+\Gamma_{S_2\to\ell\bar\ell'S_1}
+\Gamma_{S_2\to\ell\bar\ell'S_3}$,\,
where \,$\Gamma_{S_l\to\ell\bar\ell'S_k}=\mbox{\small$\sum$}_{b,d\,}
\Gamma_{S_l\to\ell_b\bar\ell_d S_k}$,\,
excluding kinematically forbidden channels.

\section{Rates of $\bm{\ell^{\prime-}\to\ell^-\ell^+\ell^-}$\label{l->lll}}

The rate of the flavor-violating decay \,$\ell^{\prime-}\to\ell^-\ell^-\ell^+$\, induced by
the $\kappa_{LR}^{}$ contribution alone from Eq.\,(\ref{ccc}) can be expressed as
\begin{eqnarray}  \label{l'2lll}
\Gamma_{\ell'\to\ell\ell\bar\ell}^{\scriptscriptstyle LR} \,\,=\,\,
\frac{|\kappa_{LR}^{}|^{4\,}\bigl|k_{\ell'\to\ell\ell\bar\ell}^{LR}\bigr|
\raisebox{1pt}{$^{2\,}$}
m_{\ell'\,}^7 m_{\ell}^2}{4096_{\;}\!\pi^3} \,,
\end{eqnarray}
where the lepton mass $m_\ell^{}$ in the final state has been neglected in the phase-space integration.
For \,$\mu^-\to e^-e^-e^+$\, and \,$\tau^-\to\mu^-\mu^-\mu^+$,\, we derive, respectively,
\begin{eqnarray}   \label{kl'2lll}
k_{\mu\to ee\bar e}^{LR} &=&
\Biggl[ \Biggl( \frac{m_{S_1}^2+m_{S_3}^2}{m_{S_1}^2-m_{S_3}^2}\ln\frac{m_{S_1}}{m_{S_3}}
- 1 \Biggr) \bigl(1-2 s_{13}^2\bigr)
- \ln\frac{m_{S_1}}{m_{S_3}} \Biggr] \frac{c_{13\,}^{}s_{13\,}^{}s_{23}}{16\pi_{\,\,}^2\Lambda^4} \,,
\nonumber \\ \nonumber \\
k_{\tau\to\mu\mu\bar\mu}^{LR} &=&
\Biggl[ \Biggl( \frac{m_{S_1}^2+m_{S_3}^2}{m_{S_1}^2-m_{S_3}^2}\ln\frac{m_{S_1}}{m_{S_3}}
- 1 \Biggr) \bigl(1-2c_{13\,}^2s_{23}^2\bigr) - \ln\frac{m_{S_1}}{m_{S_3}} \Biggr]
\frac{c_{13\,}^2 c_{23\,}^{} s_{23}}{16\pi_{\,\,}^2\Lambda^4} \,,
\end{eqnarray}
upon making the approximation \,$m_{S_1}=m_{S_2}$.\,

Similarly, the rate of \,$\ell^{\prime-}\to\ell_1^-\ell_1^-\ell_2^+$\, due to $\kappa_L^{}$
alone from Eq.\,(\ref{ccc}) is
\begin{eqnarray}  \label{Gl'2lll}
\Gamma_{\ell'\to\ell_1^-\ell_1^-\ell_2^+}^{\scriptscriptstyle L} \,\,=\,\,
\frac{|\kappa_L^{}|^{4\,}\bigl|k_{\ell'\to\ell_1\ell_1\bar\ell_2}^L\bigr|
\raisebox{1pt}{$^{2\,}$}m_{\ell'}^5}{3072_{\;}\!\pi^3} \,.
\end{eqnarray}
For \,$\mu^-\to e^-e^-e^+$\, and \,$\tau^-\to\mu^-\mu^-e^+$,\, we get, respectively,
\begin{eqnarray}
k_{\mu\to ee\bar e}^L &=& \!
\left[ \left(
\frac{\displaystyle m_{S_1}^2 m_{S_3\,}^2{\rm ln}\frac{m_{S_1}}{m_{S_3}}}{m_{S_1}^2-m_{S_3}^2}
- \frac{m_{S_1}^2+m_{S_3}^2}{4} \right) \bigl(1-2 s_{13}^2\bigr)
+ m_{S_1}^2{\rm ln}\frac{\Lambda}{m_{S_1}^{}} - m_{S_3}^2{\rm ln}\frac{\Lambda}{m_{S_3}^{}}
\right] \frac{c_{13}^{}s_{13}^{}s_{23}}{\pi_{\,\,}^2\Lambda^4} \,,
\nonumber \\ \nonumber \\
k_{\tau\to\mu\mu\bar e}^L &=&
\Biggl( \frac{m_{S_1}^2 m_{S_3}^2}{m_{S_1}^2-m_{S_3}^2}\,{\rm ln}\frac{m_{S_1}}{m_{S_3}}
- \frac{m_{S_1}^2+m_{S_3}^2}{4} \Biggr)
\frac{2c_{13\,}^3s_{13\,}^{}c_{23\,}^{}s_{23}^2}{\pi_{\,\,}^2\Lambda^4} \,.
\end{eqnarray}

\end{document}